\documentclass[journal]{IEEEtran}
\IEEEoverridecommandlockouts
\usepackage{soul,xcolor}
\usepackage{bm}
\usepackage{graphics}
\usepackage{amsmath}
\usepackage[normalem]{ulem}
\usepackage{multirow,enumitem}
\usepackage{algorithm}
\usepackage{algpseudocode}
\usepackage{verbatim}
\usepackage{amsmath}
\usepackage{amsfonts}
\usepackage{amssymb}
\usepackage{mathrsfs}
\usepackage{booktabs}
\usepackage{url}
\usepackage[pdftex]{graphicx}
\usepackage{ifthen}
\usepackage{xspace}
\usepackage{dsfont}
\usepackage{bbm}
\usepackage{cite}
\usepackage{epsfig}
\usepackage{epstopdf}
\usepackage{array}
\usepackage{multicol}
\usepackage{algpseudocode}
\usepackage{algorithm}
\usepackage{amssymb}
\usepackage{breqn}
\usepackage{esint}
\usepackage{multicol}
\usepackage{amsthm}
\usepackage{setspace}
\usepackage{lipsum}
\theoremstyle{plain}

\theoremstyle{definition}

\theoremstyle{remark}

\usepackage{footnote}
\usepackage{subcaption}
\usepackage{float}
\usepackage{graphicx}
\DeclareMathOperator*{\argmax}{\arg\!\max}
\DeclareMathOperator*{\argmin}{\arg\!\min}

\algnewcommand{\Initialize}[1]{%
  \State \textbf{initialize}
  \Statex \hspace{\algorithmicindent}\parbox[t]{.94\linewidth}{\raggedright #1}
}

\title{Protecting Legacy Wireless Systems Against Interference: Precoding and Codebook Approaches Using Massive MIMO and Region Constraints
}

\author{
\IEEEauthorblockN{Sameer Mathad, Taejoon Kim, and David J. Love}
\thanks{A preliminary version of this paper was presented at the 2024 Asilomar Conference on Signals, Systems, and Computers. This work is in part supported by the National Science Foundation (NSF) under grants EEC1941529, CNS2212565, CNS2225577, CNS2225578, and ITE2326898.}
\thanks{Sameer Mathad and David J. Love are with the Elmore Family School of Electrical and Computer Engineering,
Purdue University, West Lafayette, USA (e-mail: smathad@purdue.edu; djlove@purdue.edu).}
\thanks{Taejoon Kim is with the School of Electrical, Computer and Energy Engineering, Arizona State University, Tempe, AZ, USA (email: taejoonkim@asu.edu).}
}

\begin{document}
\bstctlcite{BSTcontrol}

\maketitle
\begin{abstract}
The ever-increasing demand for high-speed wireless communication has generated significant interest in utilizing frequency bands that are adjacent to those occupied by legacy wireless systems. Since the legacy wireless systems were designed based on often decades-old assumptions about wireless interference, utilizing these new bands will result in interference with the existing legacy users. Many of these legacy wireless devices are used by critical infrastructure networks upon which society depends. There is an urgent need to develop schemes that can protect legacy users from such interference. For many applications, legacy users are located within geographically-constrained regions. Several studies have proposed mitigating interference through the implementation of exclusion zones near these geographically-constrained regions. In contrast to solutions based on geographic exclusion zones, this paper presents a communication theory-based solution. By leveraging knowledge of these geographically-constrained regions, we aim to reduce the interference impact on legacy users. We achieve this by incorporating received power constraints, termed as region constraints, in our massive multiple-input multiple-output (MIMO) system design. We perform a capacity analysis of single-user massive MIMO and a sum-rate analysis of the multi-user massive MIMO system with transmit power and region constraints. We present a precoding design method that allows for the utilization of new frequency bands while protecting legacy users.
\end{abstract}


\section{Introduction}

\nocite{Asilomar}

The demand for high-speed wireless communication has increased rapidly over the last decade due to the proliferation of devices such as smartphones, laptops, tablets, and, more generally, wireless-enabled electronics. To meet the demand, a straightforward approach is to allocate more spectrum for broadband wireless communication \cite{Spectrum}. Unfortunately, there is a limited amount of spectrum for potentially licensed access, especially at frequencies below 6 GHz. Wireless operators have pushed for access to bands that are neighboring legacy users such as radar altimeters, Global Positioning System (GPS), etc. The need for sharing the spectrum with legacy users may potentially cause previously unexpected interference issues, e.g., when the frequency-domain sidelobes from downlink or uplink communication raise the interference level in legacy bands \cite{turf_war,gps}. These problems will likely continue into 6G and beyond systems \cite{6G}.

Massive multiple-input multiple-output (MIMO) system has become a crucial technology for meeting the demand for high-speed wireless communication \cite{Prospective,Larsson_MIMO,emil_mimo}. However, existing design and analysis of massive MIMO system do not account for emerging interference challenges and ongoing public policy discussions. The sub-6 GHz frequencies are heavily used by legacy systems critical to important networks relied upon by society, including transportation, defense, and public safety infrastructure.  Because of the growing demand for wireless communication, the Federal Communications Commission (FCC) and other spectrum authorities have identified several new sub-6 GHz bands usable by LTE, 5GNR, and Wi-Fi. 

Many, if not all, legacy spectrum allocation decisions were made decades ago, reflecting the technology of the past. Over the last decade, the main concern has been that the traditional frequencies are heavily populated and the necessary technical improvements needed are viewed as too costly \cite{Spectrum}. This has resulted in recently exacerbated and previously unforeseen interference challenges because most of the receivers deployed decades ago are unregulated \cite{interference1,Coexistence}. This means that they were designed when spectral bands were lightly used and under the assumption that no other transmitters were operating in even distant bands. The filters used in these receivers may have sidebands that allow problematic levels of interference at bands several hundreds MHz away \cite{threshold}. Even worse, the number of devices equipped with wireless technology must be deployed in close proximity in frequency, space, and time. This results in a further increase in the already problematic interference level. 

One example of such an interference scenario is 5G C-band deployment interfering with radar altimeter receivers \cite{Coexistence}. The separation between 5G C-Band and the spectrum used by radar altimeters is thought to be insufficient, primarily due to the absence of any front-end rejection requirements at the altimeter receivers. It has been speculated that this might result in interference level that could potentially corrupt the radar altimeter measurements. Because of this issue, there was a delay in the deployment of 5G systems using these bands near airports, hospitals, and other areas of use. Some studies have looked at the coexistence of radar altimeter with 5G base stations in the 4.2-4.4 GHz operating frequency \cite{alti3,alti2,alti1,Coexistence}. Ref. \cite{alti3} has measured the interference power from a 5G base station at various altitudes. Studies in \cite{alti2,Coexistence} suggest that there should be an exclusion zone around the aircraft flight path for 5G base stations operating in the C-band. The Federal Aviation Administration (FAA) has stated that permanent changes to the 5G rules should be made to ensure the safety of aircrafts \cite{aircraft}. Although establishing an exclusion zone addresses public safety concerns, it is essential to incorporate this issue into the communication system model to ensure the safety of aircraft and facilitate the coexistence of radar altimeters and 5G base stations.

Another issue is discussed in the Ligado-GPS dispute, where Ligado tried to create a wideband wireless network using the L-band \cite{gps,fcc}. The frequencies that they wanted to use were adjacent to the bands used by GPS. The GPS industry argued that re-purposing this band would lead to interference with the GPS signal \cite{gps}. Ligado later proposed a lower-power solution to address the interference problem. The much discussed public policy solutions to these interference issues are mostly infeasible solutions in the long term with the ever-increasing demand for new unused bands. Because of these issues, there is an urgent need for communication theory solutions that are capable of incorporating these legacy wireless receivers into the existing system model.

\begin{figure}
  \centering
  \includegraphics[width=.99\linewidth]{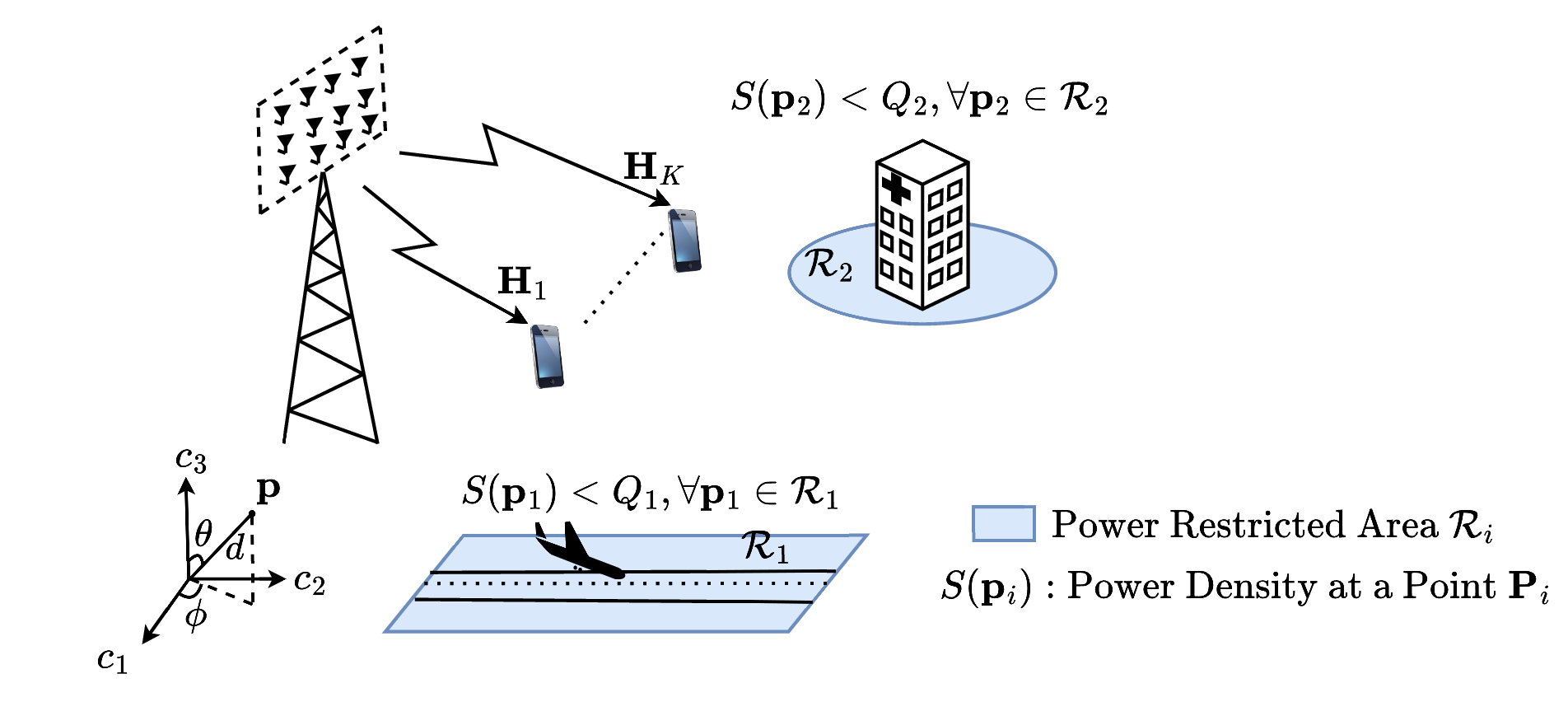}
  \caption{Power restricted regions using legacy wireless systems.}
  \label{fig:draw}
\end{figure}

Transmit precoding is now widely used in almost all wireless broadband systems. Most of the prior works on massive MIMO precoder design have not considered the interference issue with legacy systems \cite{Prospective}. In this paper, we show how to incorporate regions such as airports, hospitals, etc., where interference concerns impose received power constraints, into our system model. Fig. \ref{fig:draw} shows an example of two such regions that contain legacy wireless systems. We assume that these regions containing legacy receivers have a fixed location. We first consider a single-user massive MIMO system. We develop a precoder design framework that extends the problems outlined in \cite{waterfill} to include constraints that limit the maximum interference power at these legacy receivers. We refer to these interference power constraints as \textit{region constraints}. We then present an optimal precoder design scheme that satisfy both the transmit power and region constraints. Next, we consider a practical scenario based on limited feedback. Precoder codebook-based limited feedback techniques have been extensively studied in the literature, and it has been deployed in both 4G LTE \cite{LTE} and 5G NR \cite{5G_NR}. Examples of precoder codebook design can be found in \cite{codebook_amir, LTE, 5G_NR, CB_vasanth, kron, CB_giannakis, CB_Santipach, CB_rao, CB_Lau} and more. For a given codebook, we demonstrate how to obtain a modified codebook such that each element satisfies both the transmit power and region constraints. Next, we consider a multi-user massive MIMO system with multi-antenna users. We develop a new framework that extends the precoder design problems outlined in \cite{BD,BD2,wmmse} to include the  region constraints. We then show methods for designing precoders that satisfy both the transmit power and region constraints. Similar to the single-user scenerio, we consider a limited feedback approach and show how to obtain a modified precoder codebook that satisfies transmit power and region constraints.

As we want to limit the received power in specific regions, we show that protecting legacy users is mathematically equivalent to precoder design with quadratic constraints. Quadratically constrained optimization problems with MIMO systems have been explored in \cite{sar1,proof1,qc1,qc3,taejoon_paper_QC}. As most of the quadratically constrained problems in the literature are convex, they can be solved using standard convex optimizing techniques. In this paper we will be using standard techniques in \cite{boyd} to design precoders with quadratic constraints. 

The remainder of the paper is organized as follows. In Section \ref{SM}, we introduce a method to compute the interference power at the legacy users confined to a specific geographic region. We show that we can restrict the interference power in the region by using quadratic constraints. In Section \ref{RC}, we look at the region constraints in more detail, and we show that the region constraints can be formulated by sampling points along the boundary of the region. In Section \ref{PTP},  we analyze the capacity of single-user massive MIMO with region constraints. Similarly, in Section \ref{multi}, we analyze the sum-rate of multi-user massive MIMO with region constraints. In Section \ref{sec:sim}, we present the simulation results for our proposed precoder design for both single-user and multi-user massive MIMO. 

\subsubsection*{Notation} All uppercase boldface letters indicate matrices, while lowercase boldface letters indicate column vectors. tr(${\bf A}$), ${\bf A}^H$, and $\left |{\bf A}\right |$  are the trace, conjugate transpose, and determinant of a matrix, respectively. ${\bf A}$ = diag\{${\bf a}\}$ represents a diagonal matrix with its diagonal entries specified by the vector ${\bf a}$. ${\mathbf{I}}$ stands for the identity matrix. $\mathbf{a}_1 \otimes \mathbf{a}_2$ is the Kronecker product of two vectors. The vector $\mathop{\mathrm{vec}}(\mathbf{A})$ is obtained by stacking the columns of matrix $\mathbf{A} = [\mathbf{a}_1,\mathbf{a}_2,\dots,\mathbf{a}_M]$ as follows: $\mathop{\mathrm{vec}}(\mathbf{A}) = [\mathbf{a}^T_1,\mathbf{a}^T_2,\dots,\mathbf{a}^T_M]^T$. $\mathbb{E}[\cdot]$ denotes statistical expectation. $\mathbb{C}^M$ and $\mathbb{C}^{M\times N}$ denotes the set of all $M\times1$ complex vectors and the set of all $M\times N$ complex matrices, respectively. $\mathbf{x} \sim \mathcal{CN} (\boldsymbol{\mu},\mathbf{\Sigma})$ means $\mathbf{ x}$ is a complex Gaussian random vector with mean vector $\boldsymbol{\mu}$ and covariance matrix $\mathbf{\Sigma}$.

\section{System Model}\label{SM}

We consider a downlink multi-user massive MIMO scenario with $K$ users, $M_t$  transmit antennas at the base station and $M_r$ receive antennas at each receiver. Assuming that we are operating with a narrow band channel, the input-output relationship for the $k^{\textrm{th}}$ user can be modeled by
\begin{equation}\label{1}
    \mathbf{y}_{k} = \mathbf{H}_{k} \mathbf{x} + \mathbf{n}_{k}, \ k = 1,...,K, 
\end{equation}
where $\mathbf{y}_{k} \in \mathbb{C}^{M_r \times 1}$ is the received vector.  ${\bf n}_k \in \ \mathbb{C}^{M_r \times 1}$ is the additive noise vector whose entries are zero mean circularly symmetric independent and identically distributed (i.i.d.) complex Gaussian random variables with a variance of $\sigma^2$, i.e., ${\bf n}_k \sim$ $\mathcal{CN} ({\bf 0}, \sigma^2{\bf I})$, and ${\bf H}_k \in \mathbb{C}^{M_r \times M_t}$ is the channel matrix. We assume a precoding transmitter with the transmitted vector $\mathbf{x} \in \mathbb{C}^{M_t\times 1}$ is
\begin{equation}\label{2}
    \mathbf{x} = \sum\limits_{k = 1}^{K} {\bf F}_k \mathbf{s}_k,
\end{equation}
where ${\bf F}_k \in \mathbb{C}^{M_t \times M}$ is the $k^{\textrm{th}}$ user's linear precoder and  $\mathbf{s}_k \in \mathbb{C}^{M\times 1}$ denotes the $k^{\textrm{th}}$ user's transmit vector of $M$ data streams, where $M \leq \min\{M_r,M_t\}$, and $KM\leq M_t$. All the transmit vectors $\mathbf{s}_k$ for $k \in \{1,...,K\}$ are mutually independent and are identically distributed such that ${\bf \mathbb{E}[s}_k] = \mathbf{0}$ and  ${\bf \mathbb{E}[s}_k \mathbf{s}_k^H] = \mathbf{I}$. If $P$ is the transmit power at the base station, then in order to satisfy the transmit power constraint the precoder must be designed such that
\begin{equation}\label{3}
    \mathbb{E}[\mathbf{ x}^H\mathbf{x}] = \sum_{k = 1}^{K} \textrm{tr}({\bf F}_k^H {\bf F}_k) \leq P.
\end{equation}

Let $\mathbf{G}_k \in \mathbb{C}^{M \times M_r}$ be the $k^{\textrm{th}}$ user's linear decoder such that the estimated transmit vector is 
\begin{equation}
    \hat{\mathbf{s}}_k = \mathbf{G}_k\mathbf{ y}_{k}.
\end{equation}
The mean squared error (MSE) matrix for the $k^{\textrm{th}}$ user is  \cite{wmmse}
\begin{align}
    \mathbf{E}_k &= \mathbb{E}[(\hat{\mathbf{s}}_k - \mathbf{s}_k)(\hat{\mathbf{s}}_k - \mathbf{s}_k)^H] \\ \begin{split}
    &= \mathbf{I} + \mathbf{G}_k\mathbf{H}_k\left(\sum_{m = 1}^K\mathbf{F}_m\mathbf{F}_m^H\right)\mathbf{H}_k^H\mathbf{G}^H_k - \mathbf{G}_k\mathbf{H}_k\mathbf{F}_k \\& \quad - \mathbf{F}^H_k\mathbf{H}^H_k\mathbf{G}^H_k + \sigma^2 \mathbf{G}_k\mathbf{G}^H_k.
    \end{split}\label{6000}
\end{align}
For fixed transmit precoders, the decoder that minimizes the $\textrm{tr}(\mathbf{E}_k)$ is  \cite{wmmse}
\begin{equation} \label{7000}
    \mathbf{G}^{\star}_k = \mathbf{F}_k^H\mathbf{H}_k^H\left(\mathbf{H}_k\left(\sum_{m = 1}^K\mathbf{F}_m\mathbf{F}_m^H\right)\mathbf{H}_k^H + \sigma^2\mathbf{I}\right)^{-1}.
\end{equation}
Using this minimum mean squared error (MMSE) decoder, the corresponding MSE matrix is  \cite{wmmse}
\begin{equation}\label{8000}
    \mathbf{E}_k = \left(\mathbf{I} + \mathbf{F}^H_k\mathbf{H}^H_k\mathbf{C}^{-1}_k\mathbf{H}_k\mathbf{F}_k\right)^{-1},
\end{equation}
where $\mathbf{C}_k$ is the effective noise covariance matrix given by
\begin{equation}
    \mathbf{C}_k = \mathbf{H}_k\left(\sum_{m = 1,m \neq k}^K\mathbf{F}_m\mathbf{F}_m^H\right)\mathbf{H}_k^H +\sigma^2\mathbf{I}.
\end{equation}
The sum-rate is  \cite{wmmse}
\begin{equation}
    \mathfrak{R}_{sum} = \sum_{k = 1}^K \log\left|\mathbf{E}_k^{-1}\right|.
\end{equation}

We assume that legacy wireless users will be confined to a specific geographic region. For example, radar altimeters will be located primarily at airports. We also assume that the base station is centered at the origin. We consider a scenario with $N$ distinct geographical regions, each containing legacy wireless users. These regions can be represented by sets of points $\mathcal{R}_i \subset \mathbb{R}^3, i = 1,\dots,N$. 

The average interference power density in the $i^{\textrm{th}}$ region is constrained such that $\mathbb{E}[S({\bf p}_i,\mathbf{x})] \leq Q_i, \forall {\bf p}_i \in \mathcal{R}_i$, where $S({\bf p}_i,\mathbf{x})$ is the instantaneous power density at point ${\bf p}_i \in \mathcal{R}_i$ due to the transmitted vector $\mathbf{x}$. $Q_i$ represents the maximum allowable interference in the $i^{\textrm{th}}$ region. If the region $\mathcal{R}_i$ is in the line-of-sight of the base station, the instantaneous received power density at any point ${\bf p}_i \in \mathcal{R}_i$ can be expressed in terms of array parameters. We assume that each array element is an isotropic radiator. The instantaneous power density is \cite{array_processing}
\begin{equation}\label{4}
    S({\bf p}_i,\mathbf{x}) = \frac{1}{4\pi d_i^{\gamma}}|AF({\bf p}_i,\mathbf{x})|^2,
\end{equation}
where $AF({\bf p}_i,\mathbf{x})$ is the array factor with the array elements illuminated by the transmit vector $\mathbf{x}$, $\gamma$ is a generalized path loss exponent, $d_i$ is the distance of the point $\mathbf{p}_i$ from the base station. 

The point ${\bf p}_i$ can be represented either in Cartesian coordinate system $(c_{1,i},c_{2,i},c_{3,i})$ or polar coordinate system ($d_i,\theta_i,\phi_i$) as shown in Fig. \ref{fig:draw}. In polar coordinates, the array response vector for a general array geometry at the transmitter can be represented by $\mathbf{a}(\theta_i,\phi_i) \in \mathbb{C}^{M_t \times 1}$, where $\theta_i$ and $\phi_i$ represent the elevation and azimuth angles of departure (AoD), respectively, corresponding to the point $\mathbf{p}_i$. We define the array factor illuminated by the transmit vector $\mathbf{x}$ as
\begin{equation}\label{5}
    AF({\bf p}_i,\mathbf{x}) = \mathbf{a}(\theta_i,\phi_i)^H\mathbf{x}.
\end{equation}
 Using \eqref{5}, the instantaneous power density at point ${\bf p}_i$ can be expressed as
\begin{equation}\label{6}
    S({\bf p}_i,\mathbf{x}) = \frac{\mathbf{x}^H\mathbf{a}(\theta_i,\phi_i)\mathbf{a}(\theta_i,\phi_i)^H\mathbf{x}}{4\pi d_i^{\gamma}}.
\end{equation}
We can re-express \eqref{6} as
\begin{equation}\label{7}
    S({\bf p}_i,\mathbf{x}) = \mathbf{x}^H\mathbf{R}_i\mathbf{x},
\end{equation}
 where $\mathbf{R}_i$ is a characteristic matrix given by
\begin{equation} \label{8}
    \mathbf{R}_i \triangleq \frac{1}{4\pi d_i^{\gamma}} \mathbf{a}(\theta_i,\phi_i)\mathbf{a}(\theta_i,\phi_i)^H.
\end{equation}
The characteristic matrix is Hermitian positive semi-definite by construction. We want to constrain the average power density in the region. The expected value of \eqref{7} is 
\begin{equation}\label{9}
   \mathbb{E}[ \mathbf{ x}^H{\bf R}_i\mathbf{x} ] = \sum\limits_{k = 1}^{K} \textrm{tr}({\bf F}_k^H{\bf R}_i{\bf F}_k).
\end{equation}

The characteristic matrix in \eqref{8} is inversely proportional to $d_i^{\gamma}$. Consequently, by constraining the received power density at the boundary of the region nearest to the base station, the path loss will naturally fulfill the received power requirement throughout the remainder of the region. The $i^{\textrm{th}}$ region's boundary can be defined using a set of points $\mathcal{B}_i$ that satisfy a curve equation. The boundary $\mathcal{B}_i$ is  the set
\begin{equation}
    \mathcal{B}_i = \{(c_1,c_2,c_3) \in \mathcal{R}_i: \kappa_c(c_1,c_2,c_3) = 0\},
\end{equation}
where $\kappa_c(c_1,c_2,c_3) = 0$ is a curve equation in Cartesian coordinates. The curve can be transformed from Cartesian coordinates to polar coordinates yielding the set
\begin{equation}
    \mathcal{B}_i = \{(d,\theta,\phi) \in \mathcal{R}_i: \kappa_p(d,\theta,\phi) = 0\},
\end{equation}
where $\kappa_p(d,\theta,\phi) = 0$ is a curve equation in polar coordinates. 

To define the region constraints we consider a sampled constraint problem. By sampling the region boundary, we can represent the interference issue in the region using a finite number of constraints. The number of samples must be sufficiently high to account for the worst-case interference. Let $L_i$ be the number of discrete samples for the $i^{\textrm{th}}$ region. By obtaining $L_i$-combined samples of azimuth and elevation angles along the boundary curve $(\theta,\phi) \in \mathcal{B}_i$, closest to the base station, denoted as $(d_{i,\ell},\theta_{i,\ell},\phi_{i,\ell})$ for $\ell = 1,2,\dots, L_i$, we can effectively constrain the power density within the region $\mathcal{R}_i$ using $L_i$ quadratic constraints.

 In order to limit the interference in the region, the power density at all points ${\bf p}_i \in \mathcal{R}_i$ must remain below a specified power threshold $Q_i$. We can approximate this power density requirement in the $i^{\textrm{th}}$ region $\mathcal{R}_i$ by using $L_i$ quadratic inequality constraints. With $N$ regions having received power requirements, we generalize the quadratic inequality constraints, referred to as region constraints, as
\begin{equation}\label{10}
    \sum\limits_{k = 1}^{K}\textrm{tr}({\bf F}_k^H{\bf R}_{i,\ell}{\bf F}_k) \leq Q_i,  i=1,2,..,N, {\ell} = 1,2,..,L_i.
\end{equation}
where ${\bf R}_{i,\ell} = \frac{1}{4\pi d_{i,\ell}^{\gamma}} \mathbf{a}(\theta_{i,\ell},\phi_{i,\ell})\mathbf{a}(\theta_{i,\ell},\phi_{i,\ell})^H$, and $(d_{i,\ell},\theta_{i,\ell},\phi_{i,\ell})$ represents the $\ell^{\textrm{th}}$ sample of the $i^{\textrm{th}}$ region boundary.

This method of defining the region constraints assumes that the region lies in the line-of-sight of the base station, which is a suitable assumption for areas like airports. However, if the region is not in the line-of-sight of the base station, the region constraints can still be defined by leveraging the region's fixed location. We can measure beamforming vectors at $L_i$ sampled points along the $i^{\textrm{th}}$ region boundary. These beamforming vectors can then be used to construct $L_i$ characteristic matrices that define the region constraints.

We assume that the channel $\mathbf{ H}_k$ is perfectly known at the transmitter and the receiver. In general the transmitter only has access to an estimated channel state information (CSI). The details of the algorithm on how the transmitter gets the channel state information is beyond the scope of this paper.

\section{Region Constraints}\label{RC}

 As introduced in the previous section, the legacy wireless users will be confined to $N$ geographic regions represented by sets of points $\mathcal{R}_i \in \mathbb{R}^3, i = 1,\dots,N$. To restrict the received power density in these regions, we use multiple quadratic inequality constraints. We use a uniform planar array (UPA) at the base station. We assume that the array elements are placed in the first quadrant of the $c_2 = 0$ plane, with the first element centered at the origin. Let $M_{1}$ be the number of vertical antennas and $M_2$ be the number of horizontal antennas. The array response matrix of a UPA is  \cite{opt_array}
\begin{equation}
    {\bf A }(\theta, \phi)= \left[\begin{matrix}1 & \cdots & e^{-j(M_2-1)\psi_2}\cr e^{-j\psi_1} & \cdots & e^{-j[\psi_1+(M_2-1)\psi_2]}\cr \vdots & \ddots & \vdots \cr e^{-j(M_1-1)\psi_1} & \cdots & e^{-j[(M_1-1)\psi_1+(M_2-1)\psi_2]} \end{matrix}\right],
\end{equation}
\begin{align}
     \psi_1 &= \frac{2\pi \delta}{\nu}\cos{\theta},\\
     \psi_2 &= \frac{2\pi \delta}{\nu}\sin{\theta}\cos{\phi},
\end{align}
 where $\delta$ is distance between the antenna array elements, and $\nu$ is the wavelength. The array response vector $\mathbf{a}(\theta,\phi) = \mathop{\mathrm{vec}}({\bf A }(\theta, \phi))$. We define
 \begin{align}
     \widetilde{\mathbf{a}}(\psi_1) &= \left[1,e^{-j\psi_1},\dots,e^{-j(M_1 - 1)\psi_1}\right]^T,\\
     \widetilde{\mathbf{a}}(\psi_2) &= \left[1,e^{-j\psi_2},\dots,e^{-j(M_2 - 1)\psi_2}\right]^T.
 \end{align}
Then we can rewrite the array response vector as a Kronecker product 
\begin{equation}\label{130}
    \mathbf{a}(\theta,\phi) = \widetilde{\mathbf{a}}(\psi_2) \otimes \widetilde{\mathbf{a}}(\psi_1).
\end{equation}

Next, we show how to compute the distance between the base station and all points on the region boundary $\mathcal{B}_i$ in terms of angle $(\theta,\phi)$. Let $\mathcal{R}'_i \subset \mathbb{R}^2$ represent the orthogonal projection of all the points in the set $\mathcal{R}_i$ onto the $c_3 = 0$ plane. Let the set of points $\Xi_i \subset \mathcal{R}'_i$ represent the boundary of the 2D region in the $c_3 = 0$ plane. We assume that for every $(c_1,c_2) \in \Xi_i$, the corresponding points $(c_1,c_2,c_3) \in \mathcal{B}_i$ can be obtained by considering all the points normal to the $c_3 = 0$ plane, starting from $c_{3,\textrm{min}}$ to a maximum height $c_{3,\textrm{max}}$. Using the equation of the boundary curve in polar coordinates $(d,\phi)$, the distance between the base station and the points in $\Xi_i$ can be expressed as a function of the angle $\phi$, denoted by $d(\phi)$. Using this, we can compute the distance between the base station and all points in $\mathcal{B}_i$ in terms of the angles $(\theta, \phi)$ as
\begin{equation}\label{180}
    d(\theta,\phi) = \frac{d(\phi)}{\sin{\theta}},\ \  (\phi,\theta) \in \mathcal{B}_i.
\end{equation} 
For any point $(d,\phi) \in \Xi_i$, the span of $\theta$ in $\mathcal{B}_i$ is $\theta \in [\theta_{\textrm{min}},\theta_{\textrm{max}}]$, where $\theta_{\textrm{max}} = \arctan\left(\frac{d(\phi)}{c_{3,\textrm{min}}}\right)$ and $\theta_{\textrm{min}} = \arctan\left(\frac{d(\phi)}{c_{3,\textrm{max}}}\right)$. Next, we consider two example equations of curves representing $\Xi_i$ and show how to convert these curve from Cartesian to polar coordinates. The following analysis can be done for various shapes of the curves. Below, we give example constructions of a line segment and a circle. 
\subsubsection*{Example 1}  Region boundary defined by multiple line segments. Let $\Upsilon_i$ be  the number of line segments used to represent the 2D boundary of the $i^{\textrm{th}}$ region. Let $\Xi_i = \bigcup\limits_{\upsilon = 1}^{\Upsilon_i} \Xi_{i,\upsilon}$, where $\Xi_{i,\upsilon}$ denotes the set of points representing the $\upsilon^{\textrm{th}}$ line segment. The equation of a line segment is 
\begin{equation}\label{210}
    \omega_{1,i,\upsilon} c_1 + \omega_{2,i,\upsilon} c_2 = \omega_{3,i,\upsilon},
\end{equation}
where the line segment is the set of points $\Xi_{i,\upsilon} = \{(c_1,c_2):  c_{1,i,\upsilon,\textrm{min}}\leq c_1 \leq c_{1,i,\upsilon,\textrm{max}}, c_{2,i,\upsilon,\textrm{min}}\leq c_2 \leq c_{2,i,\upsilon,\textrm{max}}, \textrm{and} \ (c_1,c_2)\ \textrm{satisfies}\ \eqref{210}\}$. If we substitute $c_1 = d\cos(\phi)$ and $c_2 = d\sin(\phi)$, \eqref{210} leads to the distance in terms of angle $\phi$ as
\begin{equation}{\label{15}}
    d(\phi) = \frac{\omega_{3,i,\upsilon}}{\omega_{1,i,\upsilon}\cos(\phi) + \omega_{2,i,\upsilon}\sin(\phi)}, \phi \in  [\phi_{i,\upsilon,\textrm{min}}, \phi_{i,\upsilon,\textrm{max}}],
\end{equation}
where $\phi_{i,\upsilon,\textrm{min}} = \min(\Phi_{i,\upsilon})$, $\phi_{i,\upsilon,\textrm{max}} = \max(\Phi_{i,\upsilon})$, and $\Phi_{i,\upsilon}$ is a set given by $\Phi_{i,\upsilon} = \left\{\phi: \phi = \arctan{\frac{c_2}{c_1}},\ \forall (c_1,c_2) \in \Xi_{i,\upsilon} \right\}$.

\subsubsection*{Example 2}  The region boundary defined by a circle not centered at origin. The equation of a circle is 
\begin{equation}\label{16}
        (c_1-\omega_{1,i})^2 + (c_2-\omega_{2,i})^2 = \omega_{3,i}^2,
\end{equation}
where $(\omega_{1,i},\omega_{2,i})$ represents the coordinates of the center of the circle and $\omega_{3,i}$ represents the radius of the circle. The circle is a set of points $\Xi_i = \{(c_1,c_2): (c_1,c_2)\ \textrm{satisfies}\ \eqref{16}\}$. If we substitute $c_1 = d\cos(\phi)$ and $c_2 = d\sin(\phi)$, \eqref{16} leads to
\begin{equation} \label{17}
    (d\cos(\phi) - \omega_{1,i})^2 + (d\sin(\phi) - \omega_{2,i})^2 = \omega_{3,i}^2.
\end{equation}
After some algebraic manipulation we get the distance in terms of angle $\phi$ as
\begin{multline}\label{18}
     d(\phi) = (\omega_{1,i}\cos(\phi) + \omega_{2,i}\sin(\phi)) \pm \\
     \sqrt{\omega_{3,i}^2 - (\omega_{1,i}^2+\omega_{2,i}^2) + (\omega_{1,i}\cos(\phi) + \omega_{2,i}\sin(\phi))^2} \ ,\\  \phi \in  [\phi_{i,\textrm{min}}, \phi_{i,\textrm{max}}],
\end{multline}
where $\phi_{i,\textrm{min}} = \min(\Phi_i)$, $\phi_{i,\textrm{max}} = \max(\Phi_i)$, and $\Phi_i$ is a set  $\Phi_i = \left\{\phi: \phi = \arctan{\frac{c_2}{c_1}},\ \forall (c_1,c_2) \in \Xi_i\right\}$

\begin{figure}
        \centering        
        \includegraphics[width=0.85\linewidth]{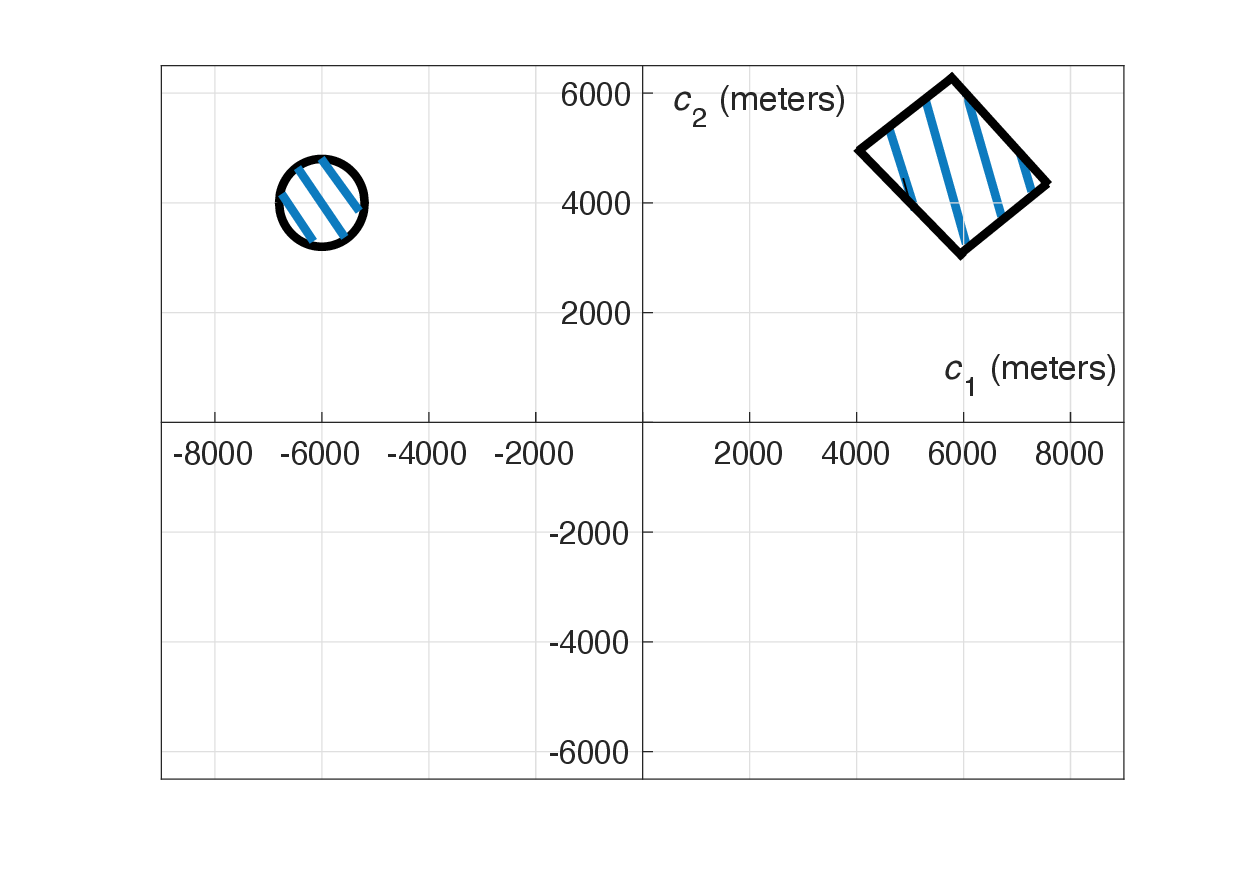}
        \caption{2D curve of region boundaries in $c_3 = 0$ plane}
        \label{fig:regions}
\end{figure}

Fig. \ref{fig:regions} displays an example of a 2D curve with two regions. One region is represented by four line segments, and the other by a circle. For the region boundary with four line segments, the line segment closest to the base station is given by the curve equation $c_2 + c_1 = 9000$, where the end points are (4000,5000) and (6000,3000). The circular region boundary is centered at (-6000,4000) with a radius of 800 meters.

\section{Capacity Analysis of Single-User massive MIMO with Region Constraint} \label{PTP}

In this section, we consider a single-user massive MIMO system with transmit power and region constraints. First, we formulate a capacity maximization problem with transmit power and region constraints, and show that the precoder structure in \cite{sar1} achieves the capacity. Next, we consider a practical limited feedback approach. We propose a method to modify any existing precoder codebook such that each element of the modified codebook satisfies both the transmit power and region constraints. Lastly, we present a sub-optimal but low-complexity solution based on an adaptive power back-off method. The single-user region constraints and transmit power constraint can be obtained from \eqref{3} and \eqref{9}, respectively, by substituting $K = 1$. The maximum achievable rate, assuming a fixed precoder $\mathbf{F}$ and i.i.d. signaling, is  \cite{waterfill}  
\begin{equation}
    C  = \log \left|\mathbf{I} + \frac{1}{\sigma^ {2}} \mathbf{H} \mathbf{F} \mathbf{F}^{H} \mathbf{H}^{H} \right|.
\end{equation}

\subsection{Capacity Maximizing Solution} \label{SU_cap}

The capacity maximizing optimization problem with region constraints and transmit power constraint is 
\begin{align*}(\mathcal{T}1)
\ \  \max_{\mathbf{F}} &\ \log \left|\mathbf{I} + \frac{1}{\sigma^ {2}} \mathbf{H} \mathbf{F} \mathbf{F}^{H} \mathbf{H}^{H} \right|,\\s.t.\ & \textrm{tr} \left(\mathbf{F}^{H} \mathbf{R}_{i,\ell}\mathbf{F}\right) \leq Q_{i}, i=1,2,..,N, \ell = 1,2,..,L_i, \\\ & \textrm{tr}\left(\mathbf{F}^{H} \mathbf{F}\right) \leq P.
\end{align*}
The Lagrangian of the problem $(\mathcal{T}1)$ is
\begin{multline}\label{24}
    \mathcal{L}(\mathbf{F}, \boldsymbol{\lambda}, \mu) = -\log \left|\mathbf{I} + \frac{1}{\sigma^ {2}} \mathbf{H} \mathbf{F} \mathbf{F}^{H} \mathbf{H}^{H} \right|+\mu \left(\textrm{tr}(\mathbf{F}^{H} \mathbf{F}) - P\right) \\
    + \sum\limits_{i=1}^{N} \sum\limits_{\ell=1}^{L_i} \lambda_{i,\ell} \left(\textrm{tr}\left(\mathbf{F}^{H} \mathbf{R}_{i,\ell} \mathbf{F}\right) - Q_{i}\right),
\end{multline}
where $\boldsymbol{\lambda} = \{\lambda_{i,\ell}\}$, and $\mu$ are, respectively, the dual variables for region constraints and transmit power constraint. For fixed dual variables the precoder structure that minimizes the Lagrangian is given by Proposition 3.1 in \cite{sar1}. We outline the optimal precoder structure. Let ${\bf Z} = \mu \mathbf{I} + \sum\limits_{i=1}^{N} \sum\limits_{\ell=1}^{L_i}  \lambda_{i,\ell} {\bf R}_{i,\ell}$. If we remove the constant term $\mu P + \sum\limits_{i=1}^{N} \sum\limits_{\ell=1}^{L_i}  \lambda_{i,\ell} Q_i$ and substitute ${\bf Z}$  in \eqref{24}, we get
\begin{equation}\label{280}
    \widetilde{\mathcal{L}}(\mathbf{F}, \boldsymbol{\lambda}, \mu) = -\log \left|\mathbf{I} + \frac{1}{\sigma^ {2}} \mathbf{H} \mathbf{F} \mathbf{F}^{H} \mathbf{H}^{H} \right|+ \textrm{tr}\left(\mathbf{F}^{H} \mathbf{Z} \mathbf{F}\right). 
\end{equation}
Suppose the singular value decomposition (SVD) of ${\bf HZ}^{-1/2}$ is
\begin{equation}{\label{4000}}
    {\bf HZ}^{-1/2} = \mathbf{U} \boldsymbol{\Gamma}^{1/2} \mathbf{V}^{H},
\end{equation}
where $\mathbf{U}$ is a $M_r \times M_r$ unitary matrix, $\mathbf{V}$ is a $M_t \times M_t$ unitary matrix, $M_t \geq M_r$, and $\boldsymbol{\Gamma}$ is a $M_r \times M_t$ real matrix with diagonal entries $\eta_1\geq \eta_2 \geq \cdots \geq \eta_{M_r} \geq 0$. $\{\eta_m\}$ are the singular values of $\mathbf{H}\mathbf{Z}^{-1/2}$. The precoder that minimizes the Lagrangian is
\begin{equation}\label{460}
    \mathbf{F}^\star_{\boldsymbol{\lambda}, \mu} = \argmin_{\mathbf{F}} \widetilde{\mathcal{L}}(\mathbf{F}, \boldsymbol{\lambda}, \mu) = \mathbf{Z}^{-1/2} \mathbf{V} \boldsymbol{\Lambda}^{1/2},
\end{equation}
where $\boldsymbol{\Lambda}$ = diag$\{\rho_1,\rho_2,\dots,\rho_{M_r}\}$ with $\rho_m = (1-\sigma^2/\eta_m)^+$ \cite{waterfill}.

The objective function of the dual problem of $(\mathcal{T}1)$ is
\begin{equation}\label{27}
g(\boldsymbol{\lambda}, \mu) =   \mathcal{L}(\mathbf{F}^\star_{\boldsymbol{\lambda}, \mu}, \boldsymbol{\lambda}, \mu).
\end{equation}
We find the optimal dual variables $\boldsymbol{\lambda}^{\star}$ and $\mu^{\star}$ that maximizes the dual problem
\begin{equation}{\label{28}}
\{\boldsymbol{\lambda}^{\star}, \mu^{\star}\} = \mathop{\arg\!\max}\limits_{\lambda_{i,\ell} \geq 0, \mu \geq 0} g(\boldsymbol{\lambda}, \mu).
\end{equation}
As the dual problem is convex, the optimal dual variables $\boldsymbol{\lambda}^{\star}$ and $\mu^{\star}$ can be found using standard convex optimization tools. The strict convexity of the primal problem $(\mathcal{T}1)$ is ensured. Thus, strong duality holds for the feasible dual problem in \eqref{27}. Therefore, $\mathbf{F}^\star_{\boldsymbol{\lambda}^{\star}, \mu^{\star}}$ is the optimal precoder. The optimal precoder waterfills over the columns of effective channel matrix ${\bf HZ}^{-1/2}$. The computational complexity of solving the dual problem is high, primarily due to the iterative method required to find the optimal dual variables. We compute the cost by counting the number of complex scalar multiplications. For example, using the sub-gradient search method described in Section \ref{sec:sim}, each iteration involves computing \eqref{460}. The matrix inverse has a cost of $O(M_t^3)$, and the SVD has a cost of $ O(M_t M_r^2)$, where $M_r \leq M_t$. The sub-gradient with respect to (w.r.t.) $\mu$ involves computing $\textrm{tr}\left(\mathbf{F}^{H} \mathbf{F}\right)$, which has a cost of $O(M_t M_r^2)$. The sub-gradient w.r.t. $\lambda_{i,\ell}$ involves computing $\textrm{tr}\left(\mathbf{F}^{H} \mathbf{R}_{i,\ell} \mathbf{F}\right)$. Since $\mathbf{R}_{i,\ell}$ is rank-one, we can write $\mathbf{R}_{i,\ell} = \mathbf{r}_{i,\ell} \mathbf{r}_{i,\ell}^H$, and thus $\textrm{tr}\left(\mathbf{F}^{H} \mathbf{R}_{i,\ell} \mathbf{F}\right) = \lVert \mathbf{F}^{H} \mathbf{r}_{i,\ell}\rVert_2^2$. Therefore, the cost of computing the sub-gradients w.r.t. $\lambda_{i,\ell}, \forall i, \ell,$ is approximately $O\left(\left(\sum\limits_{i = 1}^{N} L_i\right)M_rM_t\right)$. In a massive MIMO system, note that $M_t \gg M_r$, and the overall cost is dominated by the matrix inverse computation in \eqref{460}. Hence, the overall cost per-iteration is approximately $O(M_t^3)$.

\subsection{Modified Precoder Codebook Design for Single-User Massive MIMO with Region Constraints}\label{SU_codebook}

To address the CSI limitations and computational complexity issue of the optimal precoder in Section \ref{SU_cap}, we consider a practical limited feedback based implementation. We assume that there is a $B$-bit error-free feedback channel between the receiver and transmitter. Both the transmitter and receiver have knowledge of a precoder codebook of size $2^B$. The receiver selects the precoder from the codebook that maximizes the capacity and sends its index to the transmitter through the $B$-bit feedback link.

For any existing codebook $\mathcal{F}$, we propose a method to modify the elements of the codebook such that the transmit power and region constraints are satisfied. Based on the optimal precoder structure in Section \ref{SU_cap}, we propose a dual precoding method to generate a region constraint-aware precoder codebook. Note that dual precoding is already deployed in both LTE \cite{LTE} and 5G NR \cite{5G_NR}. We construct the codebook  $\mathcal{F}^{(P,\{Q_i\})}_{\textrm{RC}}$ which is a modified precoder codebook for a power constraint $P$ and a set of interference thresholds $\{Q_i\}$. Let $\mathbf{F}_{\zeta}$ denote the $\zeta^{\textrm{th}}$ element of codebook $\mathcal{F}$. To obtain the elements of codebook $\mathcal{F}^{(P,\{Q_i\})}_{\textrm{RC}}$, we modify the precoder $\mathbf{F}_{\zeta}$ of the codebook $\mathcal{F}$ as
\begin{equation}\label{5533}
    \mathbf{F}_{\zeta,\boldsymbol{\lambda}} = \sqrt{\frac{P}{\textrm{tr}\left(\mathbf{F}^{H}_{\zeta}\mathbf{Z}^{-1}_{\zeta,\boldsymbol{\lambda}}\mathbf{F}_{\zeta}\right)}}\mathbf{Z}^{-1/2}_{\zeta,\boldsymbol{\lambda}}\mathbf{F}_{\zeta}, \forall \mathbf{F}_{\zeta} \in \mathcal{F},
\end{equation}
where
\begin{equation}
    \mathbf{Z}_{\zeta,\boldsymbol{\lambda}} = \left(\mathbf{I} + \sum\limits_{i=1}^{N} \sum\limits_{\ell=1}^{L_i}  \lambda_{\zeta,i,\ell} {\bf R}_{i,\ell} \right),
\end{equation}
and $\lambda_{\zeta,i,\ell} \geq 0, \forall i,\ell$, where $\boldsymbol{\lambda}$ is the set of variables $\{\lambda_{\zeta,i,\ell}\}$. Our goal is to find the minimum value of $\lambda_{\zeta,i,\ell}, \forall i,\ell,$ such that the region constraints are satisfied.

Similar to the complementary slackness condition of problem $(\mathcal{T}1)$, we observe that if the $(i,\ell)^{\textrm{th}}$ region constraint is active, then $\exists\ \lambda_{\zeta,i,\ell} > 0$ such that $\textrm{tr}\left(\mathbf{F}^{H}_{\zeta,\boldsymbol{\lambda}}\mathbf{R}_{i,\ell}\mathbf{F}_{\zeta,\boldsymbol{\lambda}}\right) = Q_i$. If the constraint is inactive, i.e., $\textrm{tr}\left(\mathbf{F}^{H}_{\zeta,\boldsymbol{\lambda}}\mathbf{R}_{i,\ell}\mathbf{F}_{\zeta,\boldsymbol{\lambda}}\right) < Q_i$, then we set $\lambda_{\zeta,i,\ell} = 0$. Let $\boldsymbol{\lambda}^{\star}$ denote the optimal set $\{\lambda^{\star}_{\zeta,i,\ell}\}$ that satisfies the complementary slackness condition $\forall i, \ell$, thereby ensuring that the region constraints are satisfied. Therefore, at optimality 
\begin{equation}\label{44000}
    \sum\limits_{i=1}^{N} \sum\limits_{\ell=1}^{L_i} \lambda^{\star}_{\zeta,i,\ell}\left(\textrm{tr} \left(\mathbf{F}^{H}_{\zeta,\boldsymbol{\lambda}^{\star}} \mathbf{R}_{i,\ell}\mathbf{F}_{\zeta,\boldsymbol{\lambda}^{\star}}\right) - Q_{i}\right) = 0.
\end{equation} 
We find the optimal $\boldsymbol{\lambda}^{\star}$ by using a sub-gradient search method. Based on the achievable optimality in \eqref{44000}, we define the sub-gradient of $\sum\limits_{i=1}^{N} \sum\limits_{\ell=1}^{L_i} \lambda_{\zeta,i,\ell}\left(\textrm{tr} \left(\mathbf{F}^{H}_{\zeta,\boldsymbol{\lambda}} \mathbf{R}_{i,\ell}\mathbf{F}_{\zeta,\boldsymbol{\lambda}}\right) - Q_{i}\right)$ w.r.t. $\lambda_{\zeta,i,\ell}$ as $Q_{i} - \textrm{tr}\left(\mathbf{F}^{H}_{\zeta,\boldsymbol{\lambda}}\mathbf{R}_{i,\ell}\mathbf{F}_{\zeta,\boldsymbol{\lambda}}\right), \forall i,\ell$. We use the constant step size method described in \cite{subgradient}, along with \eqref{44000} as the stopping criteria. The details of the sub-gradient search method is described in Section \ref{sec:sim}. Therefore, given a power constraint $P$ and a set of interference thresholds $\{Q_i\}$, the optimal $\mathbf{F}_{\zeta,\boldsymbol{\lambda}^{\star}} \in \mathcal{F}^{(P,\{Q_i\})}_{\textrm{RC}}$ is obtained for each $\mathbf{F}_{\zeta} \in \mathcal{F}$. We assume that both the transmitter and receiver have knowledge of this modified codebook. The precoder for transmission is selected via a brute-force search over the codebook by maximizing
\begin{equation}
    \mathbf{F}^{\star} = \argmax\limits_{\mathbf{F}_{\zeta} \in \mathcal{F}^{(P,\{Q_i\})}_{\textrm{RC}}} \log \left|\mathbf{I} + \frac{1}{\sigma^ {2}} \mathbf{H} \mathbf{F}_{\zeta} \mathbf{F}^H_{\zeta} \mathbf{H}^{H} \right|.
\end{equation}

\subsection{Adaptive Power Back-off}\label{SU_adp}

Another sub-optimal but one-shot solution to the interference problem is \textit{adaptive power back-off} \cite{sar1}. The general idea of adaptive power back-off is to design the precoder using only the transmit power constraint and check to see if the region constraints are satisfied. If the region constraints are not satisfied, the transmit power is reduced to ensure the constraints are satisfied. The capacity maximizing precoder $\mathbf{F}_0$ is first designed based on only the transmit power constraints as outlined in \cite{waterfill}
\begin{equation}
    \mathbf{F}_0 = \argmax_{\textrm{tr}(\mathbf{F}^{H} \mathbf{F}) \leq P} \log \left|\mathbf{I} + \frac{1}{\sigma^ {2}} \mathbf{H} \mathbf{F} \mathbf{F}^{H} \mathbf{H}^{H} \right|.
\end{equation}
It is then checked to see if the region constraints are satisfied. If the constraints are not satisfied, the power is then reduced to ensure the worst-case interference is below the threshold. Let $\Delta_{i,\ell} = \textrm{tr}\left({\bf F}_{0}^{H}{\bf R}_{i,\ell}{\bf F}_{0}\right), \forall i,\ell$, then the power-reduction factor is computed as
\begin{equation}
    \alpha = \textrm{min} \left\{1,\frac{Q_1}{\Delta_{1,1}},\cdots,\frac{Q_i}{\Delta_{i,\ell}},
    \cdots,\frac{Q_{N}}{\Delta_{N,L_N}}\right\},
\end{equation}
and the precoder for transmission is ${\bf F} = \sqrt{\alpha} {\bf F}_0$.

\section{Sum-Rate Analysis of Multi-User Massive MIMO with Region Constraint} \label{multi}

In this section, we formulate the optimization problem for multi-user massive MIMO with transmit power and region constraints. We extend the joint precoder and decoder design framework in \cite{wmmse} to include region constraints. We formulate a sum-rate maximizing problem with transmit power and region constraints and provide an optimal precoder and decoder structure. Next, we extend the block-diagonalization (BD)-based precoder design scheme in \cite{BD} to include region constraints and provide the optimal solution. Then, we provide a practical solution based on limited feedback. We show how to get a modified precoder codebook that satisfy transmit power and region constraints. Lastly, we provide a sub-optimal but low-complexity solution based on power back-off method.

 \subsection{Sum-Rate Maximizing Precoder With Region Constraints}\label{MU_det}

The sum-rate maximization problem with the transmit power and region constraints is
\begin{align*}(\mathcal{T}2)
  \  &\max_{\{\mathbf{F}_1, \dots,  \mathbf{F}_K\}} \ \sum_{k = 1}^K \log\left|\mathbf{E}_k^{-1}\right|,  \\
 &\ s.t. 
 \sum_{k = 1}^K \textrm{tr} \left(\mathbf{F}^H_{k} \mathbf{R}_{i,\ell} \mathbf{F}_{k}  \right) \leq Q_i,i=1,..,N,\ell = 1,..,L_i,       
 \\
& \qquad \sum_{k = 1}^K  \textrm{tr} \left(\mathbf{F}^H_{k} \mathbf{F}_{k}  \right) \leq P. 
\end{align*}
The Lagrangian of the problem ($\mathcal{T}2$) is formulated as
\begin{multline}\label{7300}
    \mathcal{L}(\{\mathbf{F}_k\},\mu,\boldsymbol{\lambda}) = -\sum_{k = 1}^K \log\left|\mathbf{E}_k^{-1}\right| + \mu \Bigg(\sum_{k = 1}^K\textrm{tr}\left(\mathbf{F}^H_{k} \mathbf{F}_{k}\right) - \\P\Bigg)  + \sum\limits_{i=1}^{N} \sum\limits_{\ell=1}^{L_i} \lambda_{i,\ell} \left(\sum_{k = 1}^K\textrm{tr}\left(\mathbf{F}^H_{k} \mathbf{R}_{i,\ell} \mathbf{F}_{k}\right) - Q_i\right),
\end{multline}
where $\{\lambda_{i,\ell}\}$ and $\mu$ are the dual variables. For fixed precoders, the optimal MMSE decoder structure is given by \eqref{7000}. Taking the gradient of \eqref{7300} w.r.t the precoder $\mathbf{F}_k$ and setting it to 0 we get \cite{wmmse}
\begin{multline}\label{7400}    
  \left(\sum_{m = 1, m \neq k}^K\mathbf{H}_m^H\mathbf{C}^{-1}_m\mathbf{H}_m\mathbf{F}_m\mathbf{E}_m\mathbf{F}^H_m\mathbf{H}^H_m\mathbf{C}^{-1}_m\mathbf{H}_m\right)\mathbf{F}_k - \\  \mathbf{H}_k^H\mathbf{C}^{-1}_k\mathbf{H}_k\mathbf{F}_k\mathbf{E}_k + \sum\limits_{i=1}^{N} \sum\limits_{\ell=1}^{L_i}\lambda_{i,\ell}\mathbf{R}_{i,\ell}\mathbf{F}_k + \mu\mathbf{F}_k = 0.
\end{multline}
Using the matrix inversion lemma, the MMSE decoder in \eqref{7000} can be written as
\begin{equation}\label{7500}
    \mathbf{G}_k = \mathbf{E}_k\mathbf{F}^H_k\mathbf{H}^H_k\mathbf{C}_k^{-1}.
\end{equation}
Substituting \eqref{8000} and \eqref{7500} in \eqref{7400} and after some algebraic manipulation, we get the precoder structure
\begin{multline}\label{8100}
\mathbf{F}_k = \bigg(\sum_{m = 1}^K\mathbf{H}_m^H\mathbf{G}_m^H\mathbf{E}^{-1}_m\mathbf{G}_m\mathbf{H}_m + \sum\limits_{i=1}^{N} \sum\limits_{\ell=1}^{L_i}\lambda_{i,\ell}\mathbf{R}_{i,\ell} +\\ \mu\mathbf{I}\bigg)^{-1} \mathbf{H}_k^H\mathbf{G}_k^H\mathbf{E}^{-1}_k, k = 1,\dots,K,    
\end{multline}
Since the optimal decoders are a function of the precoders and vice versa, \cite{wmmse} has proposed an iterative procedure to find the solution which is outlined in Algorithm \ref{algo:3}. While the problem $(\mathcal{T}2)$ is not jointly convex with respect to both the precoder and decoder matrices, it is convex when either the decoders are designed for fixed precoders or the precoders are designed for fixed decoders. 

We extend the convergence analysis in \cite{wmmse} to include region constraints. Due to the alternating optimization procedure, the cost function of $(\mathcal{T}2)$ increases monotonically. For a given transmit power constraint and region constraints, the sum-rate is upper bounded. This guarantees that Algorithm \ref{algo:3} at least converges to a local maximum \cite{wmmse}. It is difficult to analytically characterize the number of iterations required for Algorithm \ref{algo:3} to converge. But similar to \cite{wmmse}, in Section \ref{sec:sim}, we empirically show that the algorithm converges in a few iterations. In each iteration, the optimal dual variables are computed using the sub-gradient search method in Algorithm \ref{algo:1}, which is detailed in Section \ref{sec:sim}. Note that adding region constraints does not affect the convergence of Algorithm \ref{algo:3}, but it has a significant impact on the computational complexity of each iteration. This is primarily because each iteration of Algorithm \ref{algo:3} involves an additional iterative procedure to find the optimal dual variables. Similar to Section \ref{SU_cap}, a computational complexity analysis can be performed for the iterative method used to find the optimal dual variables. In a massive MIMO system, note that $M_t \gg KM_r$, thus the overall cost is dominated by computation of $K$ matrix inversions per-iteration in \eqref{8100}, where each matrix inversion is of size $M_t\times M_t$. Therefore, the overall computational cost is approximately $O(KM_t^3)$. 
\begin{algorithm}
\caption{Iterative sum-rate maximizing solution}\label{algo:3}
\begin{algorithmic}[1]
\Initialize { $\mathbf{F}^{(0)}_k = \sqrt{\frac{P}{\textrm{tr}(\mathbf{H}_k^H\mathbf{H}_k)K}}\mathbf{H}_k^H  \forall k$, $\tau$ = 1, and let $\epsilon>0$ be the sub-optimality gap.}  
\Repeat 
\State  Compute $\left\{\mathbf{G}^{(\tau)}_k\right\}$ for given $\left\{\mathbf{F}^{(\tau-1)}_k\right\}$ using \eqref{7000}.
\State Compute $\left\{\mathbf{E}^{(\tau)}_k\right\}$ for given $\left\{\mathbf{F}^{(\tau-1)}_k\right\}$ using \eqref{8000}.
\State Compute $\left\{\mathbf{F}^{(\tau)}_k\right\}$ for given $\left\{\mathbf{G}^{(\tau)}_k\right\}$ and $\left\{\mathbf{E}^{(\tau)}_k\right\}$, using \eqref{8100} and the sub-gradient search method in Algorithm \ref{algo:1}.
\State $\tau = \tau+1$.
\Until
 $\left|\mathfrak{R}^{(\tau-1)}_{sum} - \mathfrak{R}^{(\tau)}_{sum}\right| \leq \epsilon$.
\end{algorithmic}
\end{algorithm}

\subsection{Block Diagonal Precoder}\label{BD}

In this section, we extend the BD method in \cite{BD,BD2} to incorporate the region constraints. The $k^{\textrm{th}}$ user's noise covariance matrix in \eqref{8000} with zero-forcing constraints $\mathbf{H}_k \mathbf{F}_{\tilde{k}} = 0, \forall \tilde{k} \neq k$ is $\mathbf{C}_k = \sigma^2 \mathbf{I}$. Therefore, the BD sum-rate \cite{BD} is 
\begin{equation}
    \mathfrak{R}^{\textrm{BD}}_{sum} = \sum_{k = 1}^K \log \left|\mathbf{I} +\frac{1}{\sigma^ {2}}\mathbf{H}_k  \mathbf{F}_{k} \mathbf{F}^H_{k} \mathbf{H}_k^{H} \right|.
\end{equation}
Since the multi-user interference is eliminated, this approach avoids the alternating optimization required to find the solution in Section \ref{MU_det}.

To satisfy the zero-forcing constraint $\mathbf{H}_k \mathbf{F}_{\tilde{k}} = 0, \forall \tilde{k} \neq k$, \cite{BD} states that the $k^{\textrm{th}}$ user's precoder must lie in the null space of the rest of the users' channel matrix which can be written as $\bar{\mathbf{H}}_k = \left[\mathbf{H}_1^H \dots \mathbf{H}_{k-1}^H \mathbf{H}_{k+1}^H \dots \mathbf{H}_{K}^H\right]^H$, where $\bar{\mathbf{H}}_k \in \mathbb{C}^{\widetilde{L}\times M_t}$, $\widetilde{L} = M_r(K-1)$, and $M_t > \widetilde{L}$. Suppose the SVD of $\bar{\mathbf{H}}_k$ is
\begin{equation} \label{500000}
    \bar{\mathbf{H}}_k = \bar{\mathbf{U}}_k \bar{\mathbf{\Sigma}}_k \left[\bar{\mathbf{V}}_k^{(1)} ~ \bar{\mathbf{V}}_k^{(0)}\right]^H = \bar{\mathbf{U}}_k \bar{\mathbf{\Sigma}}_k \bar{\mathbf{V}}_k^H,
\end{equation} where $\bar{\mathbf{U}}_k$ is a $\widetilde{L}\times \widetilde{L}$ unitary matrix, $\bar{\mathbf{V}}_k = \left[\bar{\mathbf{V}}_k^{(1)} ~ \bar{\mathbf{V}}_k^{(0)}\right]$ is a $M_t\times M_t$ unitary matrix, and $\bar{\mathbf{\Sigma}}_k$ is a $\widetilde{L}\times M_t$ real matrix with non-negative diagonal entries. $ \bar{\mathbf{V}}_k^{(0)}$ holds the last $(M_t$ - rank($\bar{\mathbf{H}}_k))$ right singular vectors. The precoder that satisfies the zero-forcing constraint has the following structure \cite{BD}
\begin{equation} \label{351}
    \mathbf{F}_{k} = \bar{\mathbf{V}}_k^{(0)} \widetilde{\mathbf{F}}_{k},\ k = 1,\dots,K,
\end{equation}
where $\bar{\mathbf{V}}_k^{(0)} \in \mathbb{C}^{M_t\times \widetilde{M}}$, $\widetilde{\mathbf{F}}_{k} \in \mathbb{C}^{\widetilde{M}\times M}$, and $\widetilde{M} = (M_t$ - rank($\bar{\mathbf{H}}_k))$. In order to transmit $M$ data streams per user, $\widetilde{M}\geq M$, this requires that the $\textrm{rank}(\bar{\mathbf{H}}_k)\leq M_t - M, \forall k$, or when all the users’ channel are full rank then $\widetilde{L} \leq M_t - M$.

Using the precoder structure in \eqref{351}, the zero-forcing constraint is satisfied. Therefore, the BD sum-rate maximizing problem with transmit power, and region constraints is 
\begin{align*}(\mathcal{T}3)
\ \max_{\left\{\widetilde{\mathbf{F}}_{k}\right\}} &\ \sum_{k = 1}^K \log \left|\mathbf{I} +\frac{1}{\sigma^ {2}}\mathbf{H}_k \bar{\mathbf{V}}_k^{(0)} \widetilde{\mathbf{F}}_{k} \widetilde{\mathbf{F}}^H_{k} \bar{\mathbf{V}}_k^{(0)H}\mathbf{H}_k^{H}  \right|,\\ \begin{split}
    s.t.& \  \sum_{k = 1}^K \textrm{tr} \left(\widetilde{\mathbf{F}}^H_{k} \bar{\mathbf{V}}_k^{(0)H} \mathbf{R}_{i,\ell} \bar{\mathbf{V}}_k^{(0)} \widetilde{\mathbf{F}}_{k} \right) \leq Q_i, \\
&\qquad \qquad \qquad i=1,\dots,N, \ell = 1,\dots,L_i, \end{split}
\\
&  \sum_{k = 1}^K  \textrm{tr} \left(\widetilde{\mathbf{F}}^H_{k} \bar{\mathbf{V}}_k^{(0)H}\bar{\mathbf{V}}_k^{(0)}\widetilde{\mathbf{F}}_{k}   \right) \leq P. 
\end{align*}
The Lagrangian of the problem $(\mathcal{T}3)$ is formulated as
\begin{multline}\label{30}
 \mathcal{L}\left(\left\{\widetilde{\mathbf{F}}_{k}\right\}, \boldsymbol{\lambda}, \mu\right) = -\sum_{k = 1}^K \log \bigg|\mathbf{I} + \frac{1}{\sigma^ {2}}\mathbf{H}_k \bar{\mathbf{V}}_k^{(0)} \widetilde{\mathbf{F}}_{k} \widetilde{\mathbf{F}}^H_{k} \times\\\bar{\mathbf{V}}_k^{(0)H}\mathbf{H}_k^{H} \bigg|  +
 \mu \left(\sum_{k = 1}^K\textrm{tr}\left(\widetilde{\mathbf{F}}^H_{k} \bar{\mathbf{V}}_k^{(0)H}\bar{\mathbf{V}}_k^{(0)}\widetilde{\mathbf{F}}_{k}\right) - P\right)+ \\  \sum\limits_{i=1}^{N} \sum\limits_{\ell=1}^{L_i} \lambda_{i,\ell} \left(\sum_{k = 1}^K\textrm{tr}\left(\widetilde{\mathbf{F}}^H_{k} \bar{\mathbf{V}}_k^{(0)H} \mathbf{R}_{i,\ell} \bar{\mathbf{V}}_k^{(0)} \widetilde{\mathbf{F}}_{k}\right) - Q_i\right),
\end{multline}
where $\{\lambda_{i,\ell}\}$ and $\mu\geq 0$ are the dual variables. For fixed dual variables $\mu$ and $\boldsymbol{\lambda}$, we derive a precoder structure that maximizes the Lagrangian. Let ${\bf Z} = \mu \mathbf{I} + \sum\limits_{i=1}^{N} \sum\limits_{\ell=1}^{L_i}  \lambda_{i,\ell} {\bf R}_{i,\ell}$. Removing the constant term $\mu P + \sum\limits_{i=1}^{N} \sum\limits_{\ell=1}^{L_i}  \lambda_{i,\ell} Q_i$ and substituting ${\bf Z}$ in \eqref{30} we get
\begin{multline} \label{420}
 \hat{\mathcal{L}}\left(\left\{\widetilde{\mathbf{F}}_{k}\right\}, \boldsymbol{\lambda}, \mu\right) = -\sum_{k = 1}^K \log \bigg|\mathbf{I} +  \frac{1}{\sigma^ {2}}\mathbf{H}_k \bar{\mathbf{V}}_k^{(0)} \widetilde{\mathbf{F}}_{k}\times \\\widetilde{\mathbf{F}}^H_{k} \bar{\mathbf{V}}_k^{(0)H}\mathbf{H}_k^{H} \bigg|  +
 \sum_{k = 1}^K\textrm{tr}\left(\widetilde{\mathbf{F}}^H_{k} \bar{\mathbf{V}}_k^{(0)H} {\bf Z} \bar{\mathbf{V}}_k^{(0)} \widetilde{\mathbf{F}}_{k}\right).
\end{multline}
Let $\widetilde{\mathbf{Z}}_k = \left(\bar{\mathbf{V}}_k^{(0)H}\mathbf{Z}\bar{\mathbf{V}}_k^{(0)}\right)$, then it can be observed that the Lagrangian can be divided into $K$ single-user sub-problems as follows
\begin{multline} \label{430}
 \mathcal{\widetilde{L}}\left(\widetilde{\mathbf{F}}_{k}, \boldsymbol{\lambda}, \mu\right) = -\log \left|\mathbf{I} + \frac{1}{\sigma^ {2}}\mathbf{H}_k \bar{\mathbf{V}}_k^{(0)} \widetilde{\mathbf{F}}_{k} \widetilde{\mathbf{F}}^H_{k} \bar{\mathbf{V}}_k^{(0)H}\mathbf{H}_k^{H} \right| \\ +
 \textrm{tr}\left(\widetilde{\mathbf{F}}^H_{k} \widetilde{\mathbf{Z}}_k \widetilde{\mathbf{F}}_{k}\right).
\end{multline}
This is similar to the single-user problem in \eqref{280}. Therefore to get the precoder structure that maximizes the Lagrangian, suppose the SVD of $\mathbf{H}_k\bar{\mathbf{V}}_k^{(0)}\widetilde{\mathbf{Z}}_k^{-1/2}$ is
\begin{equation}{\label{630}}
\mathbf{H}_k\bar{\mathbf{V}}_k^{(0)}\widetilde{\mathbf{Z}}_k^{-1/2} = \mathbf{U}_k \boldsymbol{\Gamma}^{1/2}_k \mathbf{V}^{H}_k,
\end{equation}
where $\mathbf{U}_k$ is a $M_r \times M_r$ unitary matrix, $\mathbf{V}_k$ is a $\widetilde{M} \times \widetilde{M}$ unitary matrix with $\widetilde{M} \geq M_r$, and $\boldsymbol{\Gamma}_k$ is a $M_r \times \widetilde{M}$ real matrix with diagonal entries $\eta_{1,k}\geq \eta_{2,k} \geq \cdots \geq \eta_{M_r,k} \geq 0$. $\{\eta_{i,k}\}$ are the singular values of $\mathbf{H}_k\bar{\mathbf{V}}_k^{(0)}\bar{\mathbf{Z}}_k^{-1/2}$. Then the precoder that minimizes the Lagrangian is
\begin{equation}\label{451}
    \mathbf{F}^\star_{k,\boldsymbol{\lambda}, \mu} = \argmin_{\widetilde{\mathbf{F}}_{k}} \mathcal{\widetilde{L}}\left(\widetilde{\mathbf{F}}_{k}, \boldsymbol{\lambda}, \mu\right) = \bar{\mathbf{V}}_k^{(0)}\widetilde{\mathbf{Z}}_k^{-1/2} \mathbf{V}_k \boldsymbol{\Lambda}_k^{1/2},
\end{equation}
where $\boldsymbol{\Lambda}_k$ = diag$\left\{\rho_{1,k},\rho_{2,k},\dots,\rho_{M_r,k}\right\}$ with $\rho_{i,k} = \left(1 - \frac{\sigma^2}{\eta_{i,k}}\right)^+$.

The objective function of the dual problem of $(\mathcal{T}3)$ is
\begin{equation}\label{33}
g(\boldsymbol{\lambda}, \mu) =   \mathcal{L}\left(\{\mathbf{F}^\star_{k,\boldsymbol{\lambda}, \mu}\}, \boldsymbol{\lambda}, \mu\right).
\end{equation}
We find the optimal dual variables $\boldsymbol{\lambda}^{\star}$ and $\mu^{\star}$ that minimizes the dual problem
\begin{equation}\label{34}
\{\boldsymbol{\lambda}^{\star}, \mu^{\star}\} = \mathop{\arg\!\max}\limits_{\lambda_{i,\ell} \geq 0, \mu \geq 0} g(\boldsymbol{\lambda}, \mu).
\end{equation}
Similar to Section \ref{SU_cap}, the dual problem is convex and the optimal dual variables can be found using standard convex optimization tools. The BD method avoids the alternating optimization in Section \ref{MU_det}, but the computational complexity related to finding the optimal dual variables remains. A complexity analysis similar to that in Section \ref{SU_cap} can be performed. However, for brevity, we provide only the overall per-iteration cost, which is approximately $O\bigg(K\bigg(\widetilde{M}^3 + \widetilde{M}^2M_t + \widetilde{M}(M_t^2 + M_r^2) + M_tM_r\bigg(M_r+\sum\limits_{i = 1}^{N} L_i\bigg)\bigg)\bigg)$.

\subsection{Modified Precoder Codebook Design for Multi-User Massive MIMO with Region Constraints}

To address the CSI limitations, similar to the Section \ref{SU_codebook}, we consider a practical limited feedback based approach. We assume that there exists a $B$-bit error-free feedback link between each user and the base station. We assume that all users and the base station have the information of the codebook of size $2^B$. The user sends the index of the selected precoder from the codebook to the base station via the $B$-bit feedback link. 

We assume that each user has equal power allocation, i.e., for a given power constraint $P$, the $k^{\textrm{th}}$ user precoder is scaled such that $\textrm{tr}\left(\mathbf{F}_k^{H} \mathbf{F}_k\right) = \frac{P}{K}$. Let the codebook $\mathcal{F}^{(P/K,\{Q_i\})}_{\textrm{RC}}$ contain the scaled elements of region constraint-aware codebook in Section \ref{SU_codebook} as $\mathbf{F}_{\zeta} = \frac{1}{\sqrt{K}} \widetilde{\mathbf{F}}_{\zeta}, \forall \widetilde{\mathbf{F}}_{\zeta} \in \mathcal{F}^{(P,\{Q_i\})}_{\textrm{RC}}$. Note that each element of $\mathcal{F}^{(P/K,\{Q_i\})}_{\textrm{RC}}$ satisfies the constraint $\textrm{tr} \left(\mathbf{F}^{H}_{\zeta} \mathbf{R}_{i,\ell}\mathbf{F}_{\zeta}\right) \leq \frac{Q_{i}}{K}, \forall i,\ell$. If we select $K$ precoders from this scaled codebook $\mathcal{F}^{(P/K,\{Q_i\})}_{\textrm{RC}}$, then both the power and region constraints are satisfied. Thus, we obtain a multi-user region constraint-aware codebook $\mathcal{F}^{(P/K,\{Q_i\})}_{\textrm{RC}}$. We assume that all the users and the base station have knowledge of this modified codebook.

The precoder for transmission is selected such that $\mathbf{F}_{k_1} \neq \mathbf{F}_{k_1}, \forall k_1 \neq k_2$. We use a greedy search method in which the precoders are selected sequentially, such that the $k^{\textrm{th}}$ precoder is
\begin{equation}
    \mathbf{F}_k = \argmax\limits_{\mathbf{F}_{\zeta} \in \mathcal{F}^{(P/K,\{Q_i\})}\setminus\mathcal{F}_{k-1}} \log \left|\mathbf{I} + \frac{1}{\sigma^ {2}} \mathbf{H}_k \mathbf{F}_{\zeta} \mathbf{F}^H_{\zeta} \mathbf{H}_k^{H} \right|,
\end{equation}
where $\mathcal{F}_{k-1}$ is the set of selected precoder for the first $k-1$ users. $\mathbf{F}_k$ is then added to the set $\mathcal{F}_{k-1}$. We assume that after the precoders for transmission are selected, all the users have the information of the selected set $\mathcal{F}_{K}$. Using $\mathcal{F}_{K}$, the optimal decoder at each user is given by \eqref{7000}.

\subsection{Adaptive Power Back-off}

Similar to single-user massive MIMO case in Section \ref{SU_adp}, an easy, but sub-optimal solution to the interference problem is adaptive power back-off. We design the precoders $ \left\{\mathbf{F}_{0,k}\right\}$ with only the transmit power constraint in \eqref{3}. $\left\{\mathbf{F}_{0,k}\right\}$ with transmit power constraint is obtained by using either the BD precoder design method outlined in \cite{BD}
\begin{equation}
    \left\{\mathbf{F}_{0,k}\right\} = \argmax\limits_{\substack{\sum\limits_{k = 1}^K\textrm{tr}\left(\mathbf{F}_{k}^H\mathbf{F}_{k}\right)\leq P,\\ \mathbf{H}_k \mathbf{F}_{\tilde{k}} = 0, \forall \tilde{k} \neq k }}\sum_{k = 1}^K \log \left|\mathbf{I} +\frac{1}{\sigma^ {2}}\mathbf{H}_k  \mathbf{F}_{k} \mathbf{F}^H_{k}\mathbf{H}_k^{H}  \right|,
\end{equation}
or the sum-rate maximizing precoder design outlined in \cite{wmmse}
\begin{equation}
   \{\mathbf{F}_{0,k}\} = \argmax\limits_{\sum\limits_{k = 1}^K\left(\mathbf{F}^H_{k} \mathbf{F}_{k}  \right) \leq P} \sum_{k = 1}^K \log \left|\mathbf{E}^{-1}_k\right|. 
\end{equation}
We then verify whether $\left\{\mathbf{F}_{0,k}\right\}$ satisfies the region constraints. If the constraints are not met, the power is reduced by a factor $\alpha$ to ensure that the worst-case interference remains below the threshold. Let $\Delta_{i,\ell} = \sum\limits_{k = 1}^K\textrm{tr}\left({\bf F}_{0,k}^{H}{\bf R}_{i,\ell}{\bf F}_{0,k}\right), \forall i.\ell$, then the power reduction factor is computed as
\begin{equation}
    \alpha = \textrm{min} \left\{1,\frac{Q_1}{\Delta_{1,1}},\cdots,
     \frac{Q_{i}}{\Delta_{i,\ell}},\cdots,\frac{Q_{N}}{\Delta_{N,L_N}}\right\}.
\end{equation}
The precoder for transmission is $\mathbf{F}_k = \sqrt{\alpha}\mathbf{F}_{0,k}, k = 1,\dots,K$.

\section{Simulation}\label{sec:sim}

In this section, we present simulation results of the proposed method for the single-user and multi-user precoder design with region constraints. We performed Monte Carlo simulations to get the average capacity for single-user massive MIMO and average sum-rate for multi-user massive MIMO. We compare the performance of the proposed optimal precoding method, the codebook-based method, the adaptive power back-off method, and precoding without the region constraints. In all simulations, the noise variance at the receiver has a fixed value of $\sigma^2 = 1$. We assume half-wavelength array element spacing, i.e., $\delta =  \nu/2$. In all simulations, the base station UPA is in a 6$\times$6 configuration, i.e., $M_1 = M_2 = 6$, and $M_t = 36$. In all the plots RC in the legend stands for region constraints.  We use a 7-bit precoder codebook for both the single-user and multi-user scenarios. A random precoder codebook is generated as described in \cite{codebook_amir}.  We then modify the elements of this random codebook such that both region constraints and transmit power constraint are satisfied as described in Section \ref{SU_codebook}. Thus, we obtain the modified codebook.

\subsection{Region Constraints}
We consider $N = 2$ regions shown in Fig. \ref{fig:regions}. Unless otherwise specified, for both regions, we take 10 equally spaced samples of the azimuth angle span, and 5 equally spaced samples of elevation angle span. Thus, the total number of samples are $L_i = 50, i = 1,2$. Using \eqref{180}, \eqref{15} and \eqref{18}, we compute the distance in terms of $(\theta_{i,\ell},\phi_{i,\ell})$ for each sampled point. We have set $c_{3,\textrm{max}} = 1500$m and $c_{3,\textrm{min}} = 0$m. The pathloss exponent for the characteristic matrix is set to $\gamma = 2$.

\subsection{Channel Model}
We use two different models for massive MIMO channel matrix $\mathbf{ H}$. First, an uncorrelated Rayleigh fading channel model, where $\mathbf{ H} \sim \mathcal{CN}(0,\mathbf{I})$. Second, a correlated geometric ray-based channel model with a single local cluster around the user to cover the line-of-sight case \cite{cluster_model,kron}. Let $\mathbf{a}(\theta_{q,t},\phi_{q,t}) \in \mathbb{C}^{M_t\times 1}$ and $\mathbf{a}(\phi_{q,r}) \in \mathbb{C}^{M_r\times 1}$ be the $q^{\textrm{th}}$ path transmit and receive array response vectors, respectively. We assume that the receiver has a uniform linear array. Thus, the receive array response vector with half-wavelength array element spacing is $\mathbf{a}(\phi) = [1,e^{-j\pi\cos{\phi}/2},\dots,e^{-j(M_r - 1)\pi\cos{\phi}/2}]^T$. Let there be $M_s$ number of scatterers near the user, then the correlated massive MIMO channel is  \cite{cluster_model,kron}
\begin{equation}\label{500}
    \mathbf{H} = \frac{1}{\sqrt{M_s}}\sum_{q = 1}^{M_s} \alpha_q \mathbf{a}(\phi_{q,r})\mathbf{a}(\theta_{q,t},\phi_{q,t})^H,
\end{equation}
where $\alpha_q \sim \mathcal{CN}(0,1)$ represents the random path loss and the random phase of the path at the receiver. $\phi_{q,t} = \bar{\phi}_{t} + \Delta \phi_{q,t}$, and $\theta_{q,t} = \bar{\theta}_{t} + \Delta \theta_{q,t}$, where $(\bar{\phi}_{t},\bar{\theta}_{t})$ is the mean azimuth and elevation AoD of the cluster. $ \Delta \phi_{q,t}$ and $\Delta \theta_{q,t}$ are the random angular perturbations distributed as $\mathcal{N}(0,\xi_1)$ and $\mathcal{N}(0,\xi_2)$, respectively. $\phi_{q,r}$ is uniformly distributed in [0,2$\pi$].

\subsection{Sub-gradient Search Method}
We outline an algorithm based on the sub-gradient search method to find the optimal dual variables for all multi-user massive MIMO precoder design methods in Section \ref{multi}. A similar approach can be used to find the dual variables for single-user precoder design. The dual variables must satisfy the following KKT conditions:
\begin{equation}
    \mu\left(\sum\limits_{k = 1}^K\textrm{tr}\left(\mathbf{F}_k^{H} \mathbf{F}_k\right) - P\right) = 0, \mu \geq 0 , \label{790}
\end{equation}
\begin{equation} 
   \lambda_{i,\ell} \left(\sum\limits_{k = 1}^K\textrm{tr}\left(\mathbf{F}^H_{k} \mathbf{R}_{i,\ell} \mathbf{F}_{k}\right) - Q_i\right) = 0, \lambda_{i,\ell} \geq 0, \forall i,\ell.\label{800}
\end{equation}

\begin{figure}
        \centering
        \includegraphics[width=0.85\linewidth]{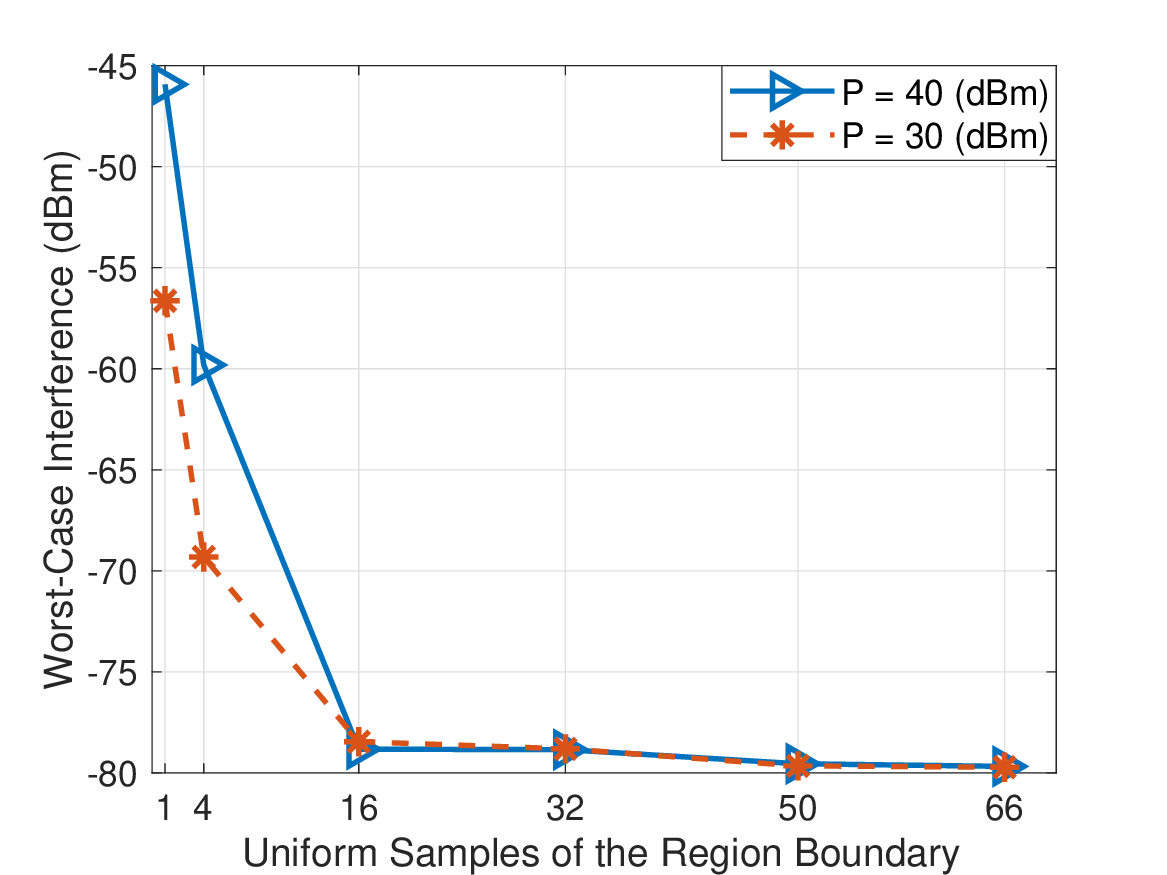}
        \caption{Worst-case interference at the region boundary vs. the number of samples of region boundary $L = L_1 = L_2$, for 36$\times$2 single-user massive MIMO and $Q_1 = Q_2 =$ -80dBm.}
        \label{fig:reg_samples}
\end{figure}

Let
\begin{multline}\label{510}
\chi = \sum\limits_{i=1}^{N} \sum\limits_{\ell=1}^{L_i} \left|\lambda_{i,\ell} \left(\sum\limits_{k = 1}^K\textrm{tr}\left(\mathbf{F}^H_{k} \mathbf{R}_{i,\ell} \mathbf{F}_{k}\right) - Q_i\right)  \right| \\+ \left| \mu\left(\sum\limits_{k = 1}^K\textrm{tr}\left(\mathbf{F}_k^{H} \mathbf{F}_k\right) - P\right) \right| ,
\end{multline}
with the achievable optimality $\chi = 0$. Using $\chi = 0$ as the stopping condition of the search algorithm, we define the sub-gradient of the dual function $g(\boldsymbol{\lambda},\mu)$ w.r.t. $\mu$ as $w_{\mu} = \left(P -\sum\limits_{k = 1}^K\textrm{tr}\left(\mathbf{F}_k^{H} \mathbf{F}_k\right)\right)$. Similarly, we define the sub-gradient of $g(\boldsymbol{\lambda},\mu)$ w.r.t. $\lambda_{i,\ell}$ as $w_{\lambda_{i,\ell}} = \left(Q_i -\sum\limits_{k = 1}^K\textrm{tr}\left(\mathbf{F}_k^{H} \mathbf{R}_{i,\ell} \mathbf{F}_k\right)\right),\forall i,\ell$. The method for solving the dual problem is outlined in Algorithm \ref{algo:1}. The step sizes $s_{\mu}$ and $s_{\lambda}$ are chosen empirically.
\begin{algorithm}
\caption{Sub-gradient search method}\label{algo:1}
\begin{algorithmic}[1]
\Initialize {$\lambda^{(0)}_{i,\ell}>0, \forall i,\ell$, $\mu^{(0)} > 0$, $\tau = 1$, and select step sizes $s_{\lambda}$ and $s_{\mu}$. Let $\epsilon>0$ be the sub-optimality gap.} 
\Repeat 
\State Given $\boldsymbol{\lambda}^{(\tau)}$ and $\mu^{(\tau)}$, compute  $\left\{\mathbf{F}^{(\tau)}_k\right\}$. 
\State  Compute $\chi^{(\tau)}$ in \eqref{510} and evaluate the sub-gradients. 
\State Update the dual variables using the constant step-size method described in \cite{subgradient}, as follows::
\State  $\lambda^{(\tau+1)}_{i,\ell} = \lambda_{i,\ell}^{(\tau)} - s_{\lambda}w_{\lambda_{i,\ell}}^{(\tau)}, \forall i,\ell$.
\State $\mu^{(\tau+1)} = \mu^{(\tau)} - s_{\mu} w_{\mu}^{(\tau)}$, $\tau = \tau+1$.
\Until
 $\chi^{(\tau)} \leq \epsilon$.
\end{algorithmic}
\end{algorithm}

\subsection{Single-user Massive MIMO}

\begin{figure}
        \centering
        \includegraphics[width=0.85\linewidth]{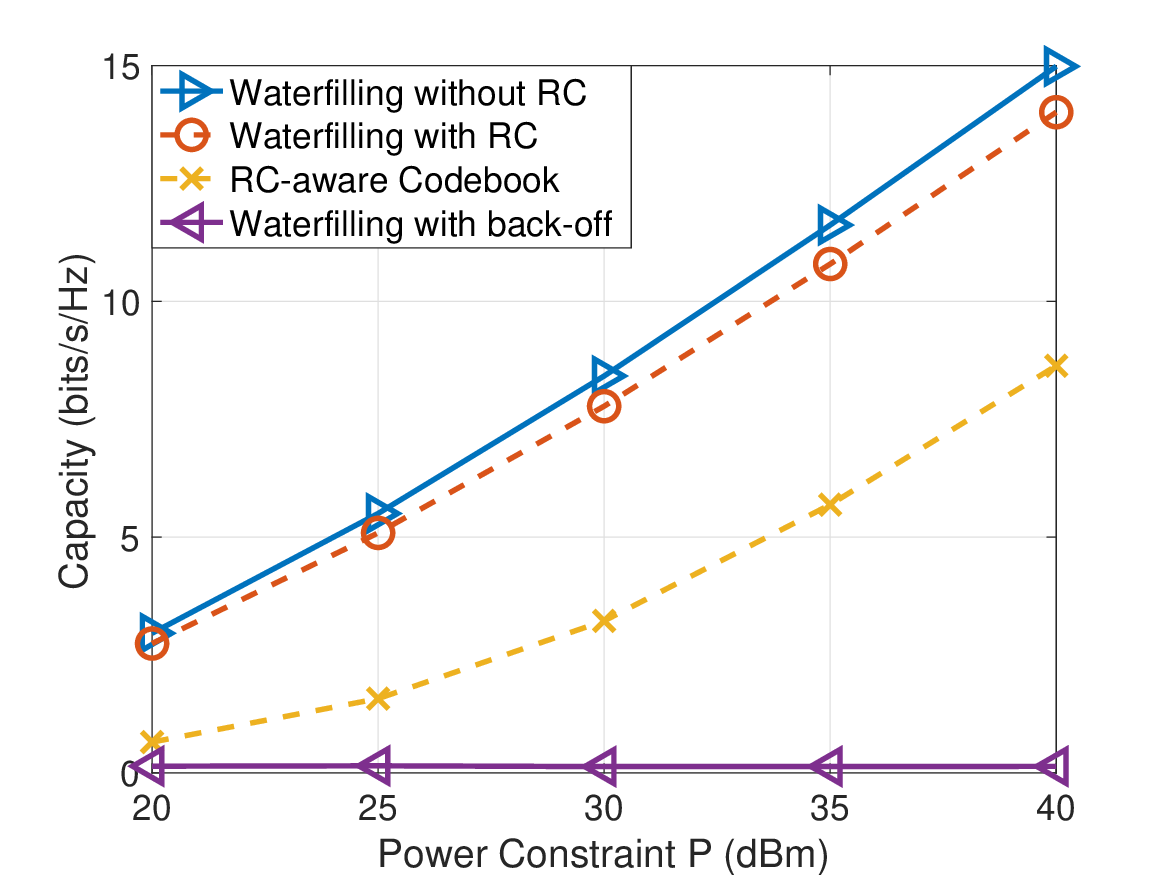}
        \caption{Power constraint P vs. capacity for 36$\times$2 single-user massive MIMO with Rayleigh fading channel.}
        \label{fig:su_mimo_p}
\end{figure}

We use the precoder design approach for the single user massive MIMO outlined in Section \ref{PTP}. In all figures, the legend ‘Waterfilling’ refers to the precoder in Section \ref{SU_cap}, while the legend 'RC-aware Codebook' refers to the precoding with the modified codebook.

Fig. \ref{fig:reg_samples} illustrates the worst-case interference at the region boundary as the number of region boundary samples $L = L_1 = L_2$ varies. We simulate the water-filling solution described in Section \ref{SU_cap}, with the interference power thresholds set to $Q_1 = Q_2 = -80$dBm. 
For each channel realization, we compute the interference at 2000 randomly selected points along the region boundary and take the maximum value as the worst-case interference. The power constraint is set to $P = 40$dBm, and $P = 30$dBm, respectively. As the boundary sample resolution increases, the gap between the main lobes of the beam steering vectors used to define the region constraints decreases. As a result, the interference in the region is more accurately captured using discrete samples along the region boundary, leading to a reduction in worst-case interference. For $L = 50$, the interference is within the required threshold. This demonstrates the necessity of incorporating region constraints in the precoder design.

\begin{figure*}
    \begin{subfigure}{0.33\textwidth}
        \centering
        \includegraphics[width=.99\linewidth]{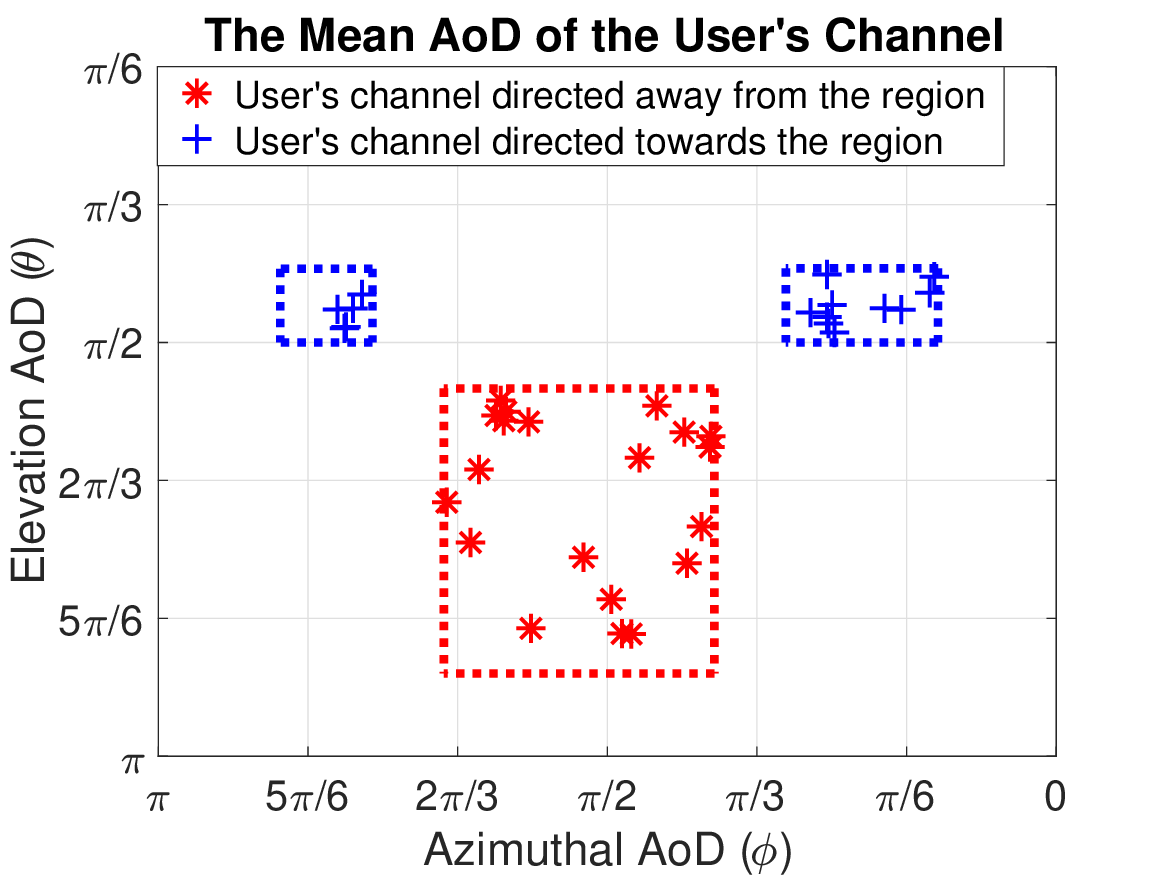}
        \caption{}
        \label{fig:dir_ch}
    \end{subfigure}
    \begin{subfigure}{0.33\textwidth}
        \centering
        \includegraphics[width=.99\linewidth]{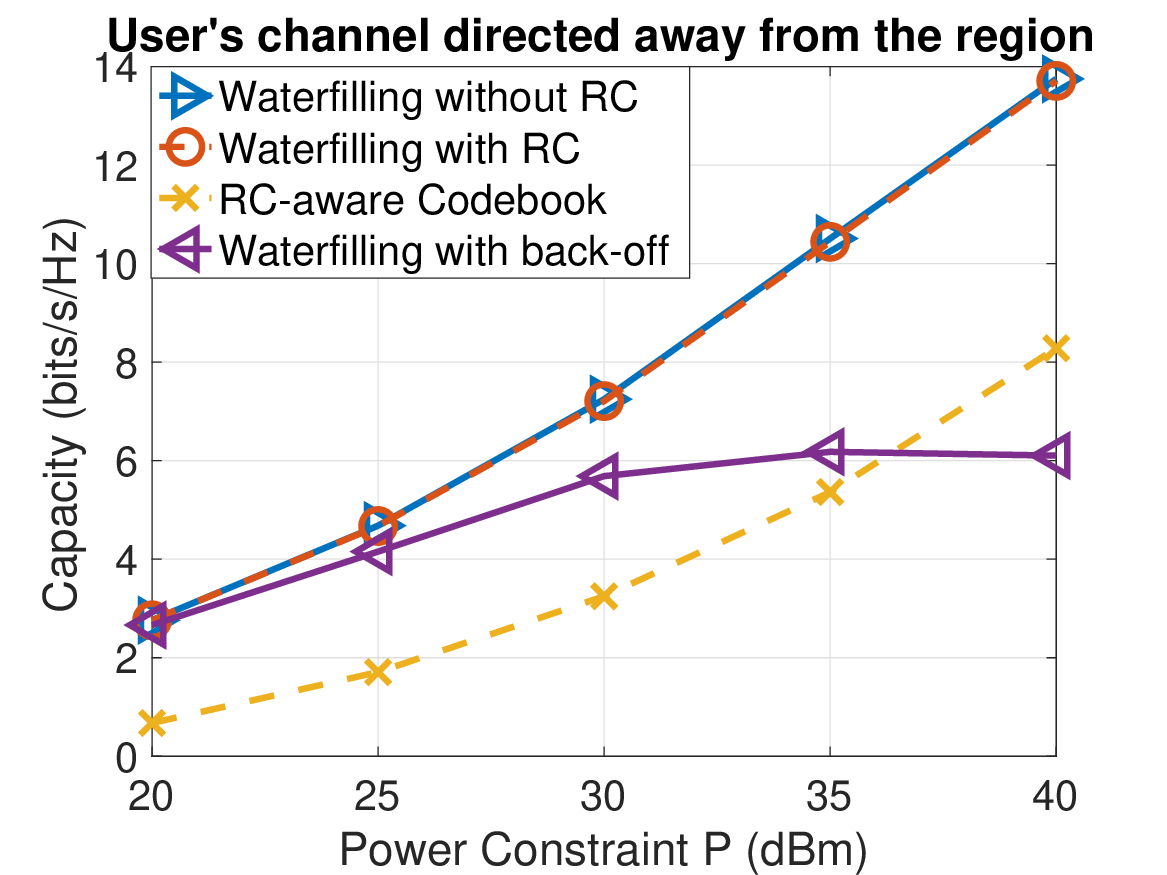}
        \caption{}
        \label{fig:su_mimo_corr1}
    \end{subfigure}
    \begin{subfigure}{0.33\textwidth}
        \centering
        \includegraphics[width=.99\linewidth]{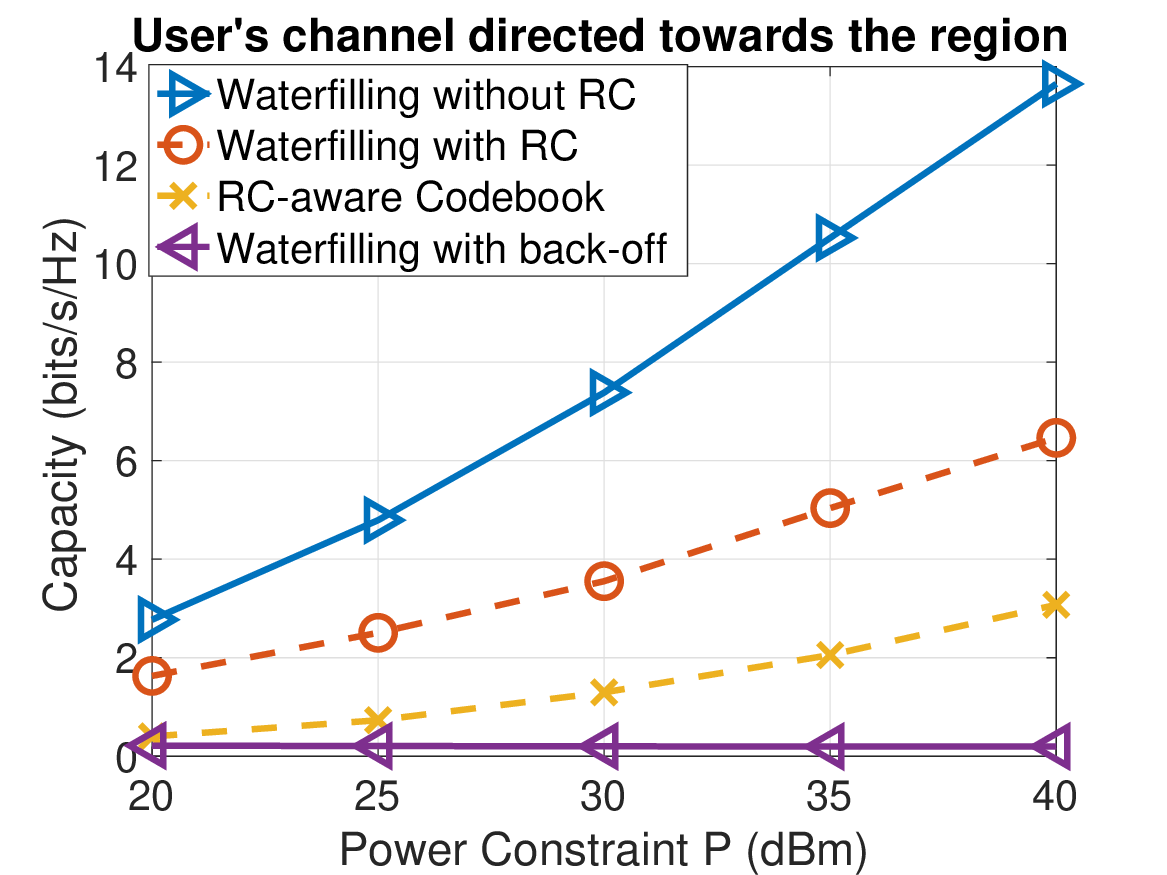}
        \caption{}
        \label{fig:su_mimo_corr2}
    \end{subfigure}
    \caption{ (a) The mean AoD of the correlated channel; (b) Power constraint P vs. capacity for user's channel directed away from the region; (c) Power constraint P vs. capacity for user's channel directed towards the region.}
\end{figure*}

\begin{figure}
        \centering
        \includegraphics[width=0.85\linewidth]{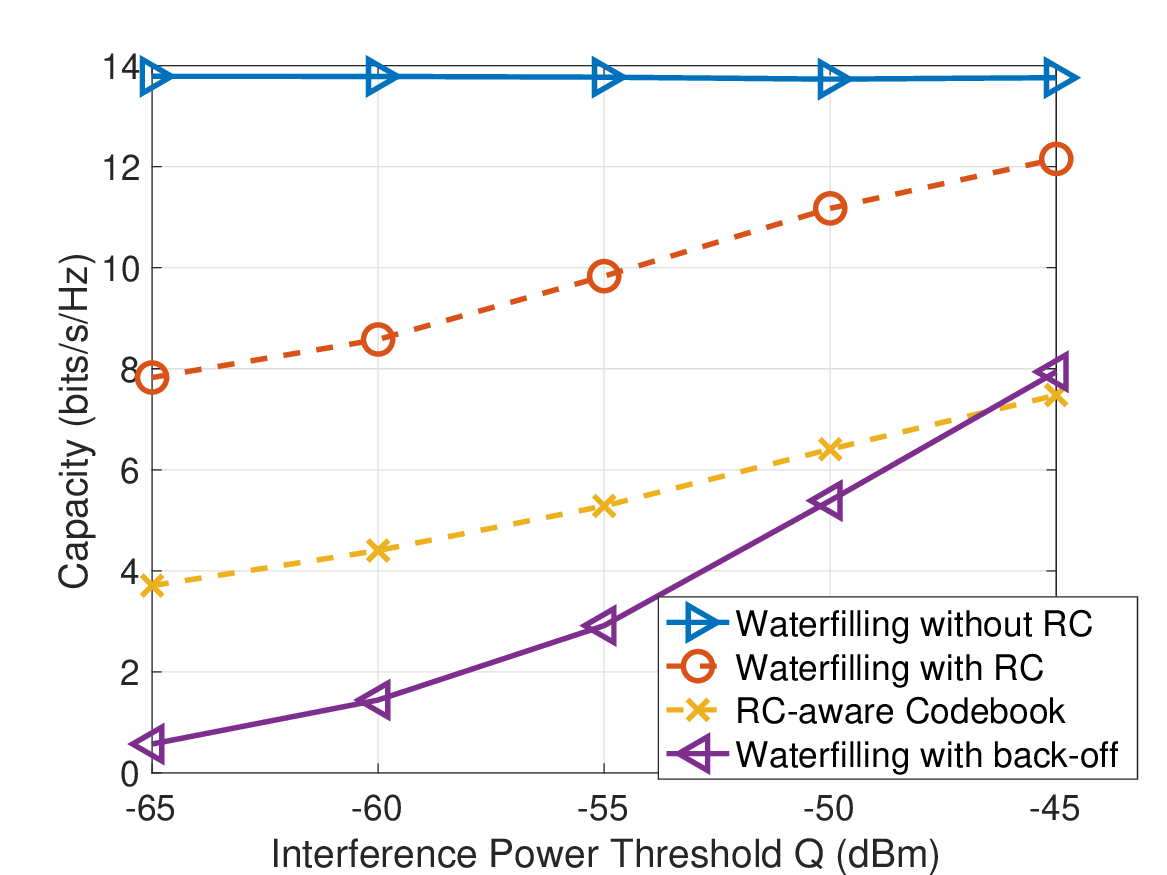}
        \caption{Interference power threshold $Q_1 = Q_2 = Q$ vs. capacity for 36$\times$2 single-user massive MIMO with user's channel directed towards the region and $P = 40$dBm.}
        \label{fig:su_mimo_q}
\end{figure}

Fig. \ref{fig:su_mimo_p} shows how the capacity varies with the power constraint $P$ for 36$\times$2 massive MIMO for Rayleigh fading channel with interference threshold $Q_1 = Q_2 = -80$dBm.  As we increase $P$, the capacity achieved by the back-off method flattens. This is because its performance is limited by the worst case interference. Where as, the capacity achieved by the proposed optimal and codebook-based methods continues to increase with $P$. For the optimal solution, the gap in the capacity with and without the region constraints increases as $P$ increases. This is because the impact of region constraints increases at higher power constraint. For example, at $P = 40$dBm, the difference in capacity is 1.03 bits/s/Hz for the waterfilling precoder. The codebook-based method outperforms the back-off method, e.g., at $P = 40$dBm, the difference in achievable capacity is 8.5 bits/s/Hz. Where as, the optimal method outperforms the codebook-based method, with capacity difference of 5.25 bits/s/Hz at $P = 40$dBm.

Fig. \ref{fig:dir_ch} illustrates a few examples of the mean AoD of the user's channel for the clustered/correlated channel model. The blue dotted boxes indicate the azimuth and elevation angle span of the regions shown in Fig. \ref{fig:regions}. Note that $\theta = 0$ corresponds to the zenith, and the azimuth axis is flipped to align with the regions shown in Fig. \ref{fig:regions}. For simulations where the user's channel is directed toward the region, the mean AoD of the clustered channel is randomly selected from within one of the two blue dotted boxes. Similarly, for simulations where the user's channel is directed away from the region, the mean AoD is randomly selected from within the red dotted box. Fig. \ref{fig:su_mimo_corr1} and Fig. \ref{fig:su_mimo_corr2} show how the capacity varies with the power constraint $P$ for a $36 \times 2$ massive MIMO system, for the two user channel choices shown in Fig. \ref{fig:dir_ch}. The interference power thresholds are $Q_1 = Q_2 = -70$dBm. The number of scatterers is $M_s = 20$, and the variances of the angular perturbation are set to $\xi_1 = 0.02$ and $\xi_2 = 0.05$.

In Fig. \ref{fig:su_mimo_corr1}, the user's channel is directed away from the region. Thus, the region constraints have minimal impact on the achievable capacity for the optimal precoding method. However, the achievable capacity for the adaptive back-off approach begins to flatten as the transmit power increases. This is because sidelobe transmission and scattering introduces interference in the region, requiring a reduction in transmit power to meet the interference threshold.

In Fig. \ref{fig:su_mimo_corr2}, the user's channel is directed towards the region. Thus, region constraints has a significant impact on the achievable capacity. For example, at $P=40$dBm, the gap in achievable capacity between the optimal method and waterfilling without region constraints is 7.19 bits/s/Hz. Similarly, the gap between the optimal and codebook-based precoding is 3.39 bits/s/Hz. Both the optimal and codebook-based precoding methods outperform the adaptive back-off method.

Fig. \ref{fig:su_mimo_q} shows how the capacity varies as we vary the interference power threshold $Q_1 = Q_2 = Q$, with power constraint $P $= 40dBm. We use the correlated channel model with the user's channel directed towards the region, as shown in Fig. \ref{fig:dir_ch}. As $Q$ increases, the difference between the capacity of the proposed waterfilling approach, waterfilling without region constraint approach, and adaptive power back-off method reduces. In the adaptive back-off method, the transmit power is scaled according to the worst-case interference. As $Q$ increases, the scale factor $\alpha$ also increases, leading to a greater increase in achievable capacity. Similarly, as $Q$ increases the capacity of the codebook-based approach also increases.

\subsection{Multi-user Massive MIMO}

\begin{figure}
        \centering
        \includegraphics[width=0.99\linewidth]{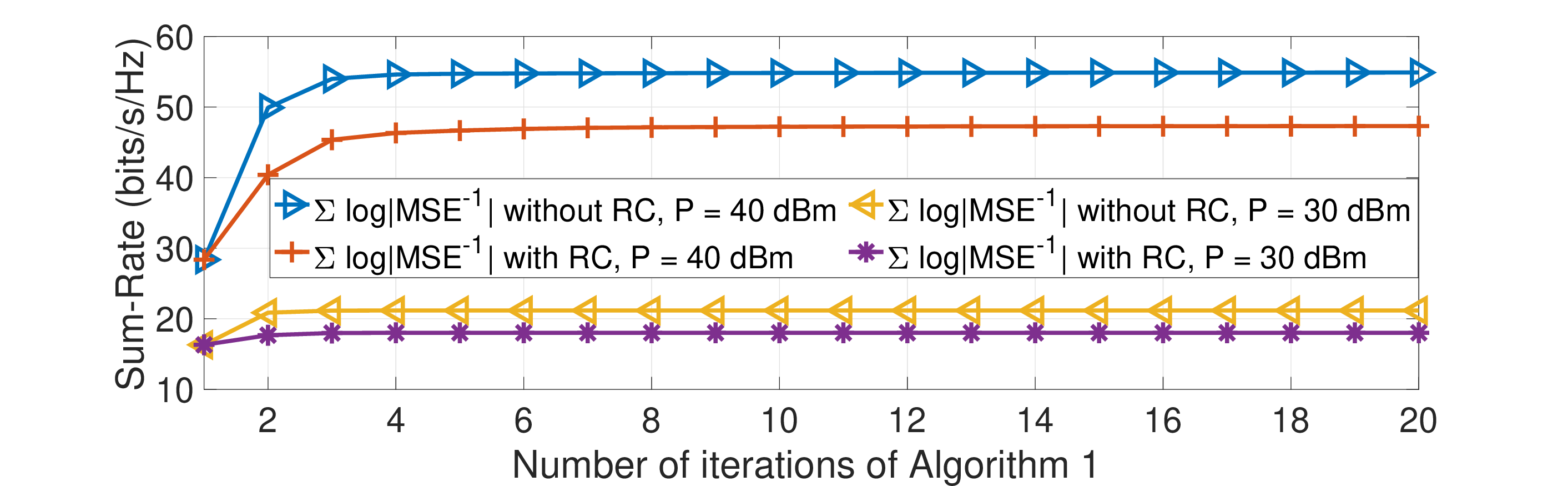}
        \caption{Convergence behavior of Algorithm \ref{algo:3} for 36$\times$2 multi-user massive MIMO with $K = 6$.}
        \label{fig:mu_mimo_cvg}
\end{figure}

\begin{figure}
        \centering
        \includegraphics[width=0.85\linewidth]{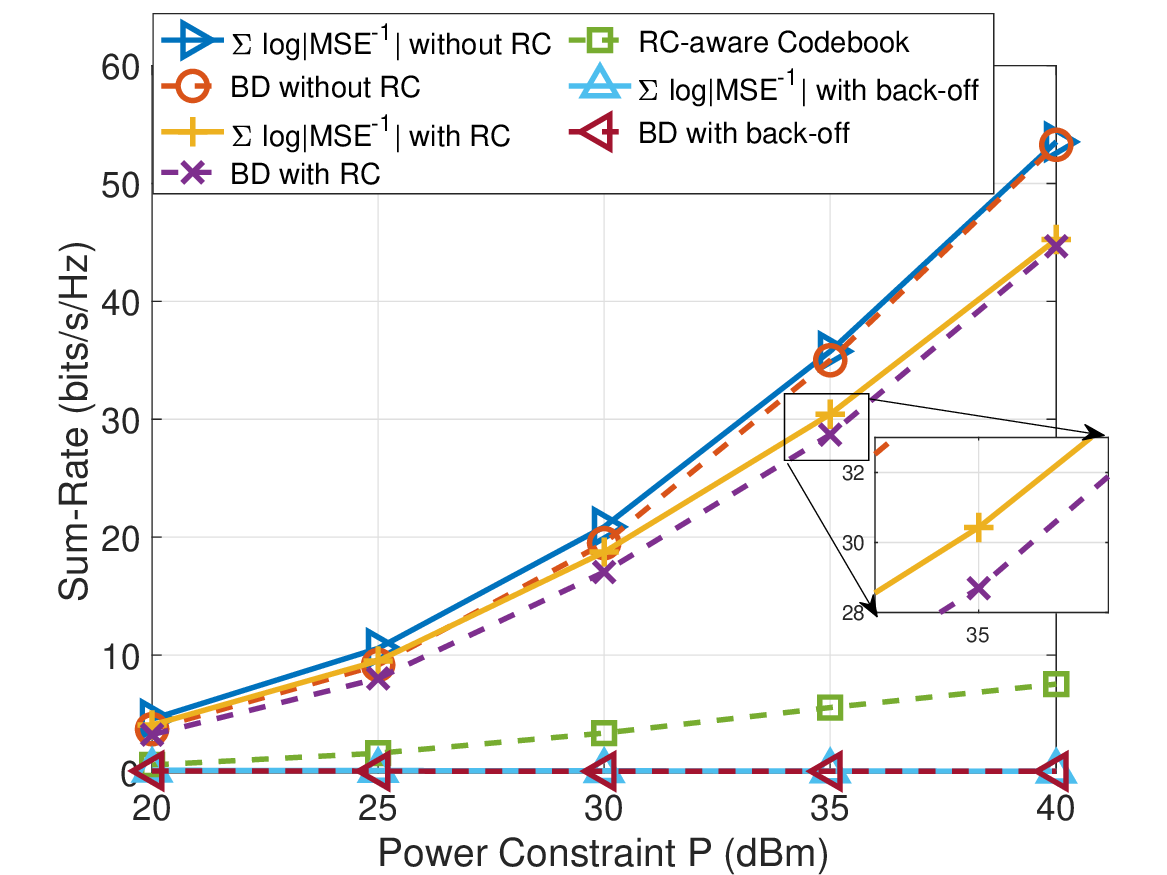}
        \caption{Power constraint P vs. sum-rate for 36$\times$2 multi-user massive MIMO with Rayleigh fading channel and $K = 6$.}
        \label{fig:mu_mimo_p}
\end{figure}

We use the precoder design approach for multi-user massive MIMO systems with region constraints as outlined in Section \ref{multi}. In all figures, the legend ‘$\sum\textrm{log}|\textrm{MSE}^{-1}|$’ refers to the precoder in Section \ref{MU_det}. Fig. \ref{fig:mu_mimo_cvg} shows the convergence behavior of the iterative sum-rate maximizing solution described in Algorithm \ref{algo:3}. The interference thresholds are $Q_1 = Q_2 = -80$dBm. We simulate a random channel realization. While the convergence speed may vary across different channel realizations, the algorithm consistently converges within a few iterations. The convergence speed is similar for both power constraints, $P = 40$dBm and $P = 30$dBm. The region constraints does not affect the convergence behavior.

Fig. \ref{fig:mu_mimo_p} shows how the sum-rate varies with $P$ for 36$\times$2 multi-user massive MIMO, $K = 6$ and a Rayleigh fading channel. The interference threshold is set at $Q_1 = Q_2 = -80$dBm. At low transmit power the $\sum\textrm{log}|\textrm{MSE}^{-1}|$ outperforms the BD precoder, for example, at $P$ = 35dBm, the difference in sum rate with region constraints active is approximately 1.75 bits/s/Hz. But at high transmit power the difference reduces, for example at $P = 40$dBm the difference is  0.4 bits/s/Hz. This is because, at low transmit power, the performance of the BD precoder is limited by noise power, whereas at high transmit power, the impact of noise becomes negligible. As $P$ increases, the sum-rate achieved by the back-off method flattens due to limitations from worst-case interference. Meanwhile, the sum-rate achieved by the proposed methods continues to increase with $P$. As $P$ increases, the gap in sum-rate with and without region constraints widens because the impact of region constraints grows. For instance, at $P$ = 40dBm, the difference in sum-rate are 8.28 bits/s/Hz, and 8.62 bits/s/Hz for the $\sum\textrm{log}|\textrm{MSE}^{-1}|$ precoder, and the BD precoder, respectively. The multi-user codebook-based precoding outperforms the back-off method, e.g., at $P = 40$dBm the difference is 7.39 bits/s/Hz.

\begin{figure}
        \centering
        \includegraphics[width=0.85\linewidth]{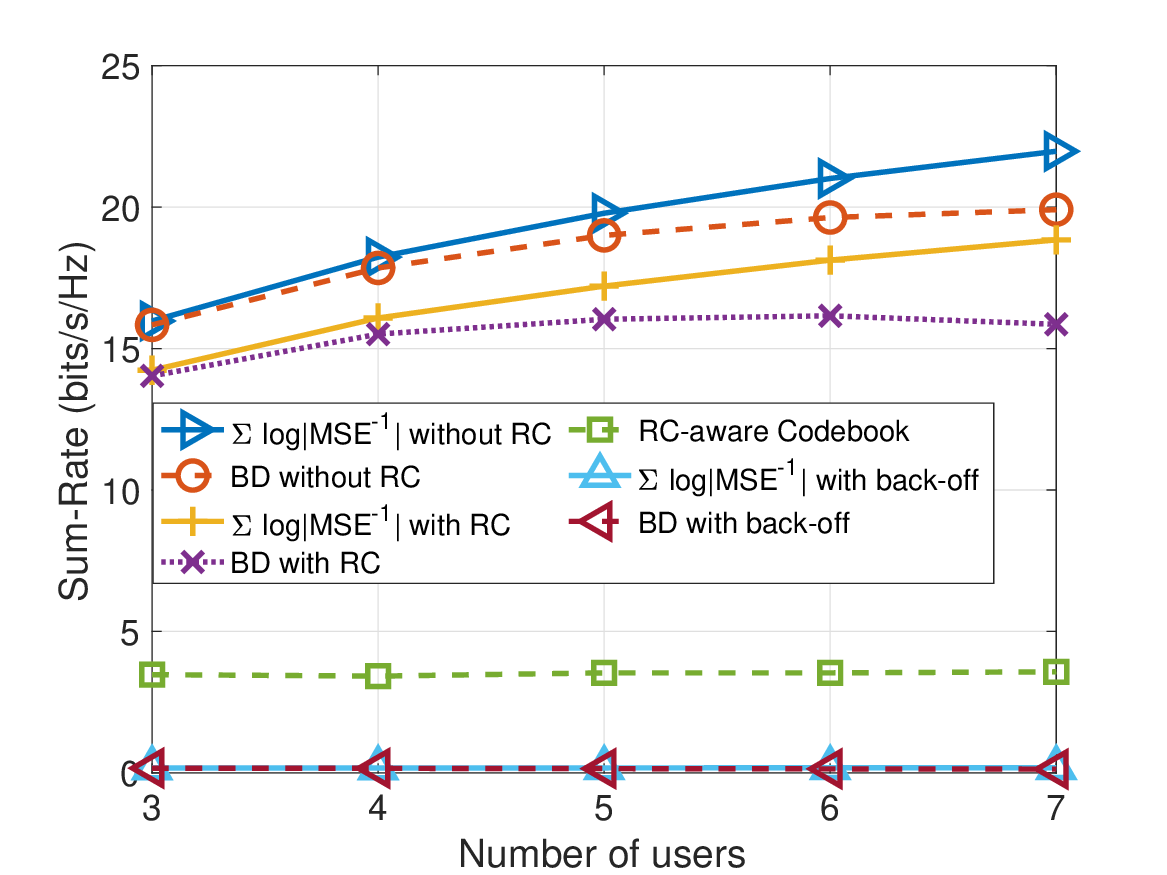}
        \caption{Number of users vs. sum-rate for 36$\times$2 multi-user massive MIMO with Rayleigh fading channel and $P = 30$dBm.}
        \label{fig:mu_mimo6}
\end{figure}

Fig. \ref{fig:mu_mimo6} shows the number of users vs. sum-rate, with $P$ = 30dBm and $Q_1 = Q_2 = -80$dBm. Similar to the previous simulation the $\sum\textrm{log}|\textrm{MSE}^{-1}|$ precoder outperforms the BD precoder. For example, with region constraints active and $K = 7$, the difference in sum-rate is approximately 3 bits/s/Hz. As $K$ increases, the sum-rate achieved by the BD precoder eventually begins to decrease. This phenomenon, previously studied in \cite{BD2}, occurs because as the number of users increases, the system loses transmit diversity to satisfy the zero-forcing constraint. The sum-rate of the codebook-based method remains unchanged as the number of users varies, since its performance is limited by the size of the codebook. 

\begin{figure}
        \centering
        \includegraphics[width=.85\linewidth]{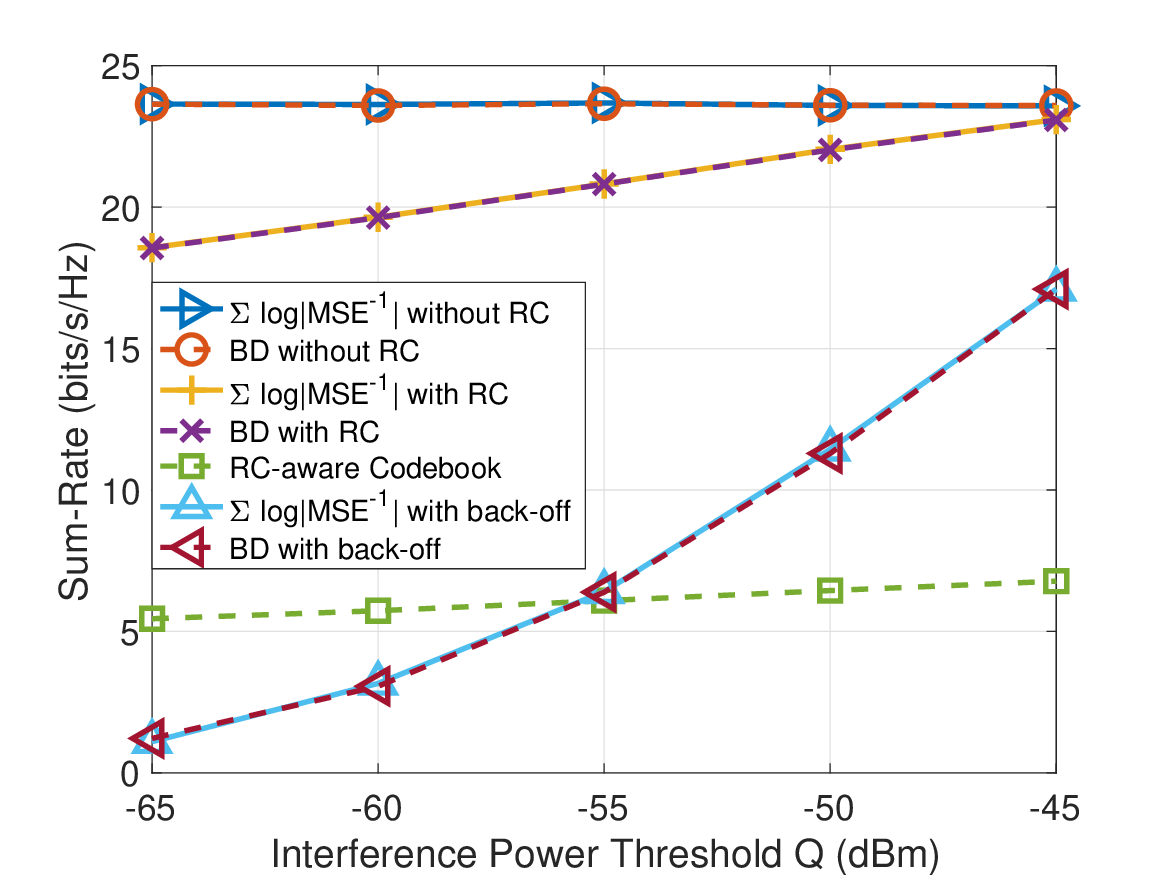}
        \caption{Interference power threshold $Q_1 = Q_2 = Q$ vs. sum-rate for 36$\times$2 multi-user massive MIMO with correlated channel, $K = 2$, and $P = 40$dBm.}
        \label{fig:MU-MIMO_n_users1}

\end{figure}

Fig. \ref{fig:MU-MIMO_n_users1} shows how the sum-rate varies as we vary the interference threshold $Q_1 = Q_2 = Q$, with $P$ = 40dBm, and $K = 2$. We use a correlated channel model, where both users channel are independent and randomly directed toward the region, as shown in Fig. \ref{fig:dir_ch}. As $Q$ increases the gap in the sum-rate between the proposed approaches, the precoding without region constraint, and adaptive back-off method reduces. In BD method, since $K = 2$, the per-user dimensionality loss due to the zero-forcing constraint is minimal, and the high transmit power constraint further reduces the impact of noise. Thus, the BD method performs similarly to $\sum\textrm{log}|\textrm{MSE}^{-1}|$. Similar to the single-user case, as $Q$ increases, the sum-rate of the codebook-based method also increases.

\section{Conclusion}
Denser utilization of the limited available spectrum has resulted in interference with legacy wireless systems. To protect the legacy users, we showed how to mathematically characterize constraints that protect the legacy users. We incorporated quadratic constraints, referred to as region constraints, into the massive MIMO system model. We formulated and solved optimization problems to develop precoders for single-user and multi-user MIMO systems. The analytical and simulation results showed the achievable capacity for single-user massive MIMO and achievable sum rate for multi-user massive MIMO systems with region constraints. For a single-user massive MIMO system, we proposed a capacity maximizing precoder design, and a practical limited feedback based method. Likewise, for multi-user massive MIMO, we proposed a sum-rate maximizing precoder design, a block-diagonalization based precoder design, and a practical limited feedback based method. The proposed precoding methods enable the use of new frequency bands while protecting the existing legacy wireless systems.

\bibliographystyle{IEEEtran}
\bibliography{IEEEabrv,ref} 

\end{document}